\documentclass[12pt]{article}
\usepackage{geometry} 
\usepackage{amsmath}
\usepackage{graphicx}

\usepackage[comma,authoryear]{natbib}

\begin{document}

\title{The trouble with orbits: the Stark effect \\ in the old and the new quantum theory}

\author{Anthony Duncan and Michel Janssen}

\maketitle

\begin{abstract}
\noindent
The old quantum theory and Schr\"odinger's wave mechanics (and other forms of quantum mechanics) give the same results for the line splittings in the first-order Stark effect in hydrogen, the leading terms in the splitting of the spectral lines emitted by a hydrogen atom in an external electric field.  We examine the account of the effect in the old quantum theory, which was hailed as a major success of that theory, from the point of view of wave mechanics. First, we show how  the new quantum mechanics solves a fundamental problem one runs into in the old quantum theory with the Stark effect. It turns out that, even without an external field, it depends on the coordinates in which the quantum conditions are imposed which electron orbits are allowed in a hydrogen atom. The allowed energy levels and hence the line splittings are independent of the coordinates used but the size and eccentricity of the orbits are not. In the new quantum theory, this worrisome non-uniqueness of orbits turns into the perfectly innocuous non-uniqueness of bases in Hilbert space. Second,  we review how the so-called WKB (Wentzel-Kramers-Brillouin) approximation method for solving the Schr\"odinger equation reproduces the quantum conditions of the old quantum theory amended by some additional half-integer terms. These extra terms remove the need for some arbitrary extra restrictions on the allowed orbits that the old quantum theory required over and above the basic quantum conditions.\footnote{We dedicate this paper to the memory of Philip M.\ Stehle (1919--2013).}
\end{abstract}




\section{Introduction}

In March of 1916, \citet{Epstein 1916a, Epstein 1916b} and \citet{Schwarzschild 1916} showed that the old quantum theory of \citet{Bohr 1913} and \citet{Sommerfeld 1915a, Sommerfeld 1915b, Sommerfeld 1916} can account for an effect discovered by and named after \citet{Stark 1913}, the splitting of the spectral lines of hydrogen atoms placed an external electric 
field.\footnote{For an account of the developments leading up to this result, see \citet{Duncan and Janssen 2014}, which, in turn, draws on \citet{Kragh 2012} and especially \citet{Eckert 2013}.} This result was hailed as a tremendous success for the old quantum theory. \citet[p.\ 150]{Epstein 1916a} boasted that ``the reported results prove the correctness of Bohr's atomic model with such striking evidence that even our conservative colleagues cannot deny its cogency." In the conclusion of the first edition of his {\it Atombau und Spektrallinien}, the bible of the old quantum theory, \citet[p.\ 458]{Sommerfeld 1919} called the theory's explanation of the Stark effect one of ``the most impressive achievements in our field" and a ``capstone on the edifice of atomic physics." 

However, as we noted in an earlier paper \citep{Duncan and Janssen 2014}, the old quantum theory's explanation of the Stark effect was not without its share of problems. These problems were solved when, shortly after the arrival of Schr\"odinger's (1926a) wave mechanics, \citet{Schroedinger 1926b} and \citet{Epstein 1926} produced an account of the Stark effect in the new theory. In this paper, we focus on two of these problems and show how they are resolved in wave mechanics. 

First, we show how  the new quantum mechanics takes care of a fundamental problem one runs into when applying the old quantum theory to the Stark effect. It turns out that the allowed orbits of the electron in the hydrogen atom, with or without an external field, depend on the coordinates in which the quantum conditions are imposed. The allowed energy levels and hence the line splittings do not depend on the coordinates used but the size and eccentricity of the elliptical orbits do. In the new quantum theory, this worrisome non-uniqueness of orbits turns into the perfectly innocuous non-uniqueness of bases in Hilbert space. 

Second,  we show how wave mechanics does away with another problem in the old quantum theory, namely the need for extra restrictions on the allowed orbits over and above the basic quantum conditions. We review how the so-called WKB approximation method for solving the Schr\"odinger equation, named after \citet{Wentzel 1926b}, \citet{Kramers 1926} and \citet{Brillouin 1926}, reproduces the quantum conditions of the old quantum theory amended by some additional half-integer terms. With these additional terms, there is no need anymore for extra restrictions on the allowed orbits. 

We will proceed as follows. In section 2, we use the old quantum theory to find the formula for the energy levels for the first-order Stark effect, i.e., the energy of the allowed orbits of an electron in a hydrogen atom placed in a weak electric field, to first order in the strength of that field. We show that the old quantum theory calls for some arbitrary restrictions on the allowed orbits over and above the basic quantum conditions. In section 3, we present the problem of the non-uniqueness of the orbits in the old quantum theory. In section 4, we sketch how the formula for the energy levels in the Stark effect is derived in wave mechanics. We only present the first part of this derivation in detail. This suffices to show how wave mechanics avoids the need for extra restrictions on the allowed quantum states. In section 5, we show how the problem of the non-uniqueness of the orbits is solved in modern quantum mechanics. In section 6, we use the WKB method to find approximate solutions to the Schr\"odinger equation to recover the quantum conditions of the old quantum theory from wave mechanics with correction terms of $\frac{1}{2}$. These correction terms remove the need for extra restrictions on the allowed orbits in the old quantum theory.  In section 7, we summarize our conclusions.

\section{The Stark effect in the old quantum theory}

In Cartesian coordinates $(x, y, z)$, the Hamiltonian for an electron (reduced mass $\mu$, charge $-e$) in a hydrogen atom in an external electric field ${\mathcal E}$ in the $z$-direction is given by (in Gaussian units): 
\begin{equation}
H = \frac{p^2}{2 \mu} - \frac{e^2}{r} + e{\mathcal E}z,
\label{Hamiltonian0}
\end{equation}
where $p^2 \equiv p_x^2 + p_y^2 + p_z^2$, with $(p_x, p_y, p_z)$ the momenta conjugate to the coordinates $(x, y, z)$. 
\begin{figure}[h]
   \centering
   \includegraphics[width=3in]{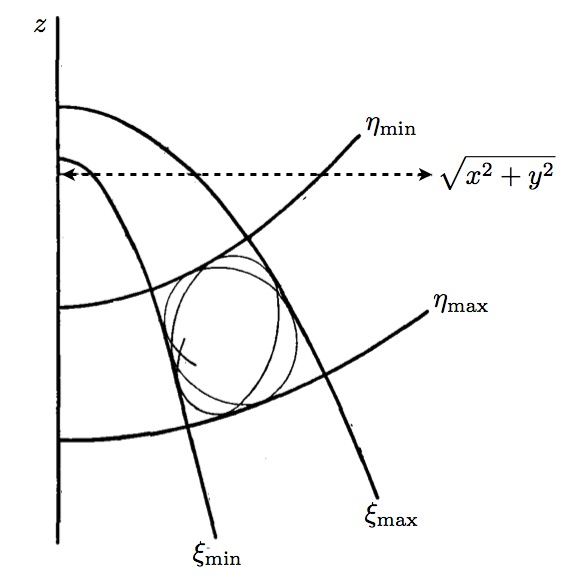} 
   \caption{{\small Parabolic coordinates. This figure is taken from \citet[p.\ 498]{Epstein 1916b} but the labeling has been changed to reflect the definition of the coordinate transformation $(x, y, z) \longrightarrow (\xi, \eta, \varphi)$ as given in Eq.\ (\ref{parabolic coordinates}), which follows  \citet[p.\ 301, Eq.\ 43]{Kramers 1919} rather than \citet[p.\ 495, Eqs.\ 19--20]{Epstein 1916b}. The figure shows what Epstein calls a ``meridian plane" (ibid.), a plane through the $z$-axis and the position of the electron. This plane rotates around the $z$-axis as the electron orbits the nucleus. Within this plane, the electron stays between $\xi_{\rm min}$ and $\xi_{\rm max}$ and between $\eta_{\rm min}$ and $\eta_{\rm max}$.}}
   \label{parabolic}
\end{figure}

We switch to parabolic coordinates $(\xi, \eta, \varphi)$, related to $(x, y, z)$ via \citep[p.\ 301, Eq.\ 43]{Kramers 1919}\footnote{\citet[p.\ 492, Eqs.\ 19--20]{Epstein 1916b} defined parabolic coordinates slightly differently. In the notation of Eq.\ (\ref{parabolic coordinates}), he set $z = (\xi^2 - \eta^2)/2$ and $\sqrt{x^2 + y^2} = \xi \eta$. Moreover, Epstein called $x$ what we call $z$ and $y$ what we call $\sqrt{x^2 + y^2}$. The quantity $r \equiv \sqrt{x^2 + y^2}$ in Epstein's notation is thus equal to $r \equiv  \sqrt{x^2 + y^2 + z^2}$ in our notation. Instead of Eq.\ (\ref{r}) below, Epstein found $r = (\xi^2 + \eta^2)/2$.}
\begin{equation}
z = \frac{\xi - \eta}{2}, \quad x+ iy = \sqrt{\xi \eta} e^{i \varphi}.
\label{parabolic coordinates}
\end{equation}
This coordinate transformation is illustrated in Fig.\ \ref{parabolic}. It follows from Eq.\ (\ref{parabolic coordinates}) that
\begin{equation}
r^2 = x^2 + y^2 + z^2 = \xi \eta + \frac{(\xi^2 - 2 \xi \eta + \eta^2)}{4} =  \frac{(\xi + \eta)^2}{4},
\end{equation}
or 
\begin{equation}
r = \frac{\xi + \eta}{2}.
\label{r}
\end{equation}

In parabolic coordinates the Hamiltonian in Eq.\ (\ref{Hamiltonian0}) is given by:
\begin{equation}
H = \frac{1}{2\mu} \left( \frac{4}{\xi + \eta} (p_\xi \xi p_\xi) +  \frac{4}{\xi + \eta} (p_\eta \eta p_\eta) + \frac{1}{\xi \eta} p^2_\varphi  \right)
- \frac{2e^2}{\xi + \eta} +  \frac{1}{2} e{\mathcal E}(\xi - \eta),
\label{Hamiltonian}
\end{equation}
where $(p_\xi, p_\eta, p_\varphi)$ are the momenta conjugate to $(\xi, \eta, \varphi)$. In the old quantum theory, as in classical mechanics, $p_\xi \xi p_\xi = \xi p_\xi^2$ and $p_\eta \eta p_\eta = \eta p_\eta^2$. The reason we wrote these products the way we did in Eq.\ (\ref{Hamiltonian}) is that in wave mechanics $p_\xi$ becomes a differential operator, differentiation with respect to $\xi$, that does not commute with multiplication by $\xi$.

Using $E$ to denote energy, we write
\begin{equation}
H = E = \alpha_1,
\label{H=alpha1}
\end{equation}
where $\alpha_1$ is some negative constant. Substituting $\alpha_1$ for $H$ in Eq.\ (\ref{Hamiltonian}), multiplying both sides of by $2\mu (\xi + \eta)$, and making the substitutions
\begin{equation}
p_\xi \longrightarrow \frac{\partial S}{\partial \xi}, \;\;\; p_\eta \longrightarrow \frac{\partial S}{\partial \eta}, \;\;\; p_\varphi \longrightarrow \frac{\partial S}{\partial \varphi},
\label{p to dS/dq}
\end{equation}
related to a canonical transformation generated by an as yet unknown function $S$, known as {\it Hamilton's principal function} \citep[p.\ 431]{Goldstein}, we obtain the Hamilton-Jacobi equation for this system in parabolic coordinates in the following form:\footnote{For a detailed explanation of the rationale behind this recipe, see, e.g., \citet[Ch.\ 10]{Goldstein} or \citet[Ch.\ 11]{Corben and Stehle}.}
\begin{equation}
4 \xi \left( \frac{\partial S}{\partial \xi} \right)^2 + 4 \eta \left( \frac{\partial S}{\partial \eta} \right)^2 + \left( \frac{1}{\xi} + \frac{1}{\eta} \right) \left( \frac{\partial S}{\partial \varphi} \right)^2 - 4\mu e^2 +  \mu e{\mathcal E}(\xi^2 - \eta^2) = 2\mu (\xi + \eta) \alpha_1.
\label{HJ eq}
\end{equation}
where we used that
\begin{equation}
\frac{\xi + \eta}{\xi \eta} = \frac{1}{\xi} + \frac{1}{\eta}, \;\;\; (\xi + \eta)(\xi - \eta) = \xi^2 - \eta^2.
\label{xi & eta}
\end{equation}
The reason for using parabolic coordinates now becomes clear. The Hamilton-Jacobi equation (\ref{HJ eq}) is {\it separable} in these coordinates. This means that its solution has the form:
\begin{equation}
S(\xi, \eta, \varphi) = S_\xi(\xi) + S_\eta(\eta) + S_\varphi(\varphi).
\label{separation}
\end{equation}
If we insert this expression for $S$ into Eq.\ (\ref{HJ eq}) and multiply both sides by $\xi \eta/(\xi + \eta)$, the only term involving $\varphi$ in the resulting equation is $(d S_\varphi/d \varphi)^2$. It follows that $d S_\varphi/d \varphi$ must be a constant, which we will call $\alpha_3$. If Eq.\ (\ref{separation}) for $S$ is inserted into Eq.\ (\ref{HJ eq}), with $d S_\varphi/d \varphi = \alpha_3$, the Hamilton-Jacobi equation splits into a part that depends only on $\xi$ and a part that depends only on $\eta$. Since the sum of these two parts must vanish, the two parts themselves must be equal but opposite constants. Writing the constant as $2 \alpha_2$, we arrive at: 
\begin{eqnarray}
 4 \, \xi \! \left( \frac{d S_\xi}{d \xi} \right)^2 + \frac{\alpha_3^2}{\xi}   - 2 \mu  e^2  + \mu e {\mathcal E} \, \xi^2 - 2 \mu \alpha_1 \xi = - 2 \alpha_2, \label{HJ xi}\\
 4 \, \eta \! \left( \frac{d S_\eta}{d \eta} \right)^2 + \frac{\alpha_3^2}{\eta}   - 2 \mu  e^2  - \mu e {\mathcal E} \, \eta^2  - 2 \mu \alpha_1 \eta = + 2 \alpha_2. \label{HJ eta}
\end{eqnarray}
The constants $\alpha_1$, $\alpha_2$, and $\alpha_3$ are called the {\it separation constants} \citep[p.\ 445]{Goldstein}. It follows from Eqs.\ (\ref{HJ xi})--(\ref{HJ eta}) that
\begin{eqnarray}
\frac{d S_\xi}{d \xi} = \frac{1}{2 \xi} \sqrt{- \alpha_3^2 + 2 (\mu  e^2 - \alpha_2) \xi + 2 \mu \alpha_1 \xi^2 - \mu e {\mathcal E} \xi^3}, \label{HJ eq xi}\\
\frac{d S_\eta}{d\eta} = \frac{1}{2 \eta} \sqrt{- \alpha_3^2 + 2 (\mu  e^2 + \alpha_2) \eta + 2 \mu \alpha_1 \eta^2 + \mu e {\mathcal E} \eta^3}. \label{HJ eq eta}
\end{eqnarray}

The quantum conditions, 
\begin{equation}
\oint p_i dq_i = n_i h
\label{quantum conditions}
\end{equation}
(where integration is over one period of the motion, the $n_i$'s are  integers, and $h$ is Planck's constant), must be imposed in coordinates in which the Hamilton-Jacobi equation is separable.\footnote{It was Schwarzschild who, in a letter of March 1, 1916, alerted Sommerfeld to the connection between the phase integral quantum conditions (\ref{quantum conditions}) introduced by \citet{Sommerfeld 1915a, Sommerfeld 1915b} and elements of celestial mechanics, especially action-angle variables and the Hamilton-Jacobi equation \citep[pp.\ 44--45; also discussed briefly in Duncan and Janssen, 2014]{Eckert 2013}.} Introducing the action variables $I_\xi$, $I_\eta$, and $I_\varphi$, we thus have:
\begin{eqnarray}
I_\xi = \oint p_\xi \, d\xi =  \oint \frac{d S_\xi}{d \xi} \, d\xi = n_\xi h, \label{I-xi}\\
I_\eta = \oint p_\eta \, d\eta = \oint \frac{d S_\eta}{d \eta} \, d\eta = n_\eta h, \label{I-eta} \\
I_\varphi = \oint p_\varphi \, d\varphi = \oint \frac{d S_\varphi}{d \varphi} \, d\varphi = n_\varphi h. \label{I-phi}
\end{eqnarray}
The action variables are so-called {\it adiabatic invariants}, which means that their values in the presence of a small electric field ${\mathcal E}$ are the same as their values in the absence of such a field.\footnote{See the passage following Eq.\ (25) in \citet{Duncan and Janssen 2014} for a brief discussion of how Ehrenfest and his student Burgers connected  quantum conditions such as those in Eqs.\ (\ref{I-xi})--(\ref{I-phi}) to the notion of adiabatic invariants and for references to the historical literature on this topic.} It is this property of  action variables in  classical mechanics that makes them suitable candidates for the quantities subjected to quantum conditions  in the old quantum theory. Where the cases ${\mathcal E}=0$ and ${\mathcal E} \neq 0$ differ is in how the separation constants depend on the action variables. As we have seen, one of these separation constants, $\alpha_1$, is just the energy $E$. So even though the action variables have the same values for ${\mathcal E}=0$ and ${\mathcal E} \neq 0$, the energy  does not. Put differently, when the electric field is switched on, the energy of the allowed orbits must change for the action variables to remain the same.

We now evaluate the integrals in Eqs.\ (\ref{I-xi})--(\ref{I-phi}) using Eqs.\ (\ref{HJ eq xi})--(\ref{HJ eq eta}) and the simple equation $d S_\varphi/d \varphi = \alpha_3$ for the integrands. This will give us expressions for the action variables $I_\xi$, $I_\eta$, and $I_\varphi$ in terms of the separation constants $\alpha_1$, $\alpha_2$, and $\alpha_3$. We then invert these relations to find the separation constants in terms of the action variables and thereby in terms of the  quantum numbers $n_\xi$, $n_\eta$, and $n_\varphi$. First, we do this for the case that ${\mathcal E}=0$, then for the case that ${\mathcal E} \neq 0$.

As long as there is no external field (${\mathcal E} = 0$), we can use the standard integral,
\begin{equation}
\int_a^b \frac{1}{x} \sqrt{(x - a)(b - x)} = \frac{\pi}{2} \left(a + b - 2 \sqrt{ab} \right),
\label{standard integral 1}
\end{equation}
to evaluate the integrals in Eqs.\ (\ref{I-xi}) and (\ref{I-eta}).

We first consider $I_\xi$. Using Eq.\ (\ref{HJ eq xi}) with ${\mathcal E}=0$ for $d S_\xi/d \xi$, we can write Eq.\ (\ref{I-xi}) as:
\begin{equation}
I_\xi =  \sqrt{- 2 \mu \alpha_1} \oint  \frac{d\xi }{2 \xi} \sqrt{ \frac{\alpha_3^2}{2 \mu \alpha_1} - \frac{\mu  e^2 - \alpha_2}{\mu \alpha_1} \xi - \xi^2}. 
\label{I_xi-0 loop integral}
\end{equation}
The quadratic expression in $\xi$ under the square root sign can be written as 
\begin{equation}
(\xi - \xi_{\rm min})(\xi_{\rm max} - \xi) = - \xi_{\rm min}\xi_{\rm max} + (\xi_{\rm min} + \xi_{\rm max}) \xi - \xi^2,
\label{roots xi}
\end{equation}
where $\xi_{\rm min}$ and $\xi_{\rm max}$ are the roots of the quadratic equation that we obtain by setting the expression under the square root sign in Eq.\ (\ref{I_xi-0 loop integral}) equal to zero. Comparing this expression to the right-hand side of Eq.\ (\ref{roots xi}), we find
\begin{equation}
\xi_{\rm min} + \xi_{\rm max} = \frac{\alpha_2 - \mu e^2}{\mu \alpha_1}, \;\;\; \xi_{\rm min}\xi_{\rm max} = -\frac{\alpha_3^2}{2 \mu \alpha_1}.
\label{a & b}
\end{equation}
Replacing the loop integral in Eq.\ (\ref{I_xi-0 loop integral}) by twice the line integral from $\xi_{\rm min}$ to $\xi_{\rm max}$, we can rewrite $I_\xi$ as
\begin{equation}
I_\xi = \sqrt{- 2 \mu \alpha_1} \int_{\xi_{\rm min}}^{\xi_{\rm max}} \frac{d\xi }{\xi} \sqrt{(\xi -\xi_{\rm min})(\xi_{\rm max} - \xi)}.
\label{I_xi-0 line integral}
\end{equation}
Using Eqs.\  (\ref{standard integral 1}) and (\ref{a & b}), we then find that
\begin{equation}
I_\xi = \frac{\pi}{2} \sqrt{- 2 \mu \alpha_1} \left[ \frac{\alpha_2 - \mu e^2}{\mu \alpha_1} -\frac{2 \alpha_3}{\sqrt{- 2 \mu \alpha_1}} \right].
\label{I_xi-0}
\end{equation}

We now turn to $I_\eta$. Using Eq.\ (\ref{HJ eq eta}) with ${\mathcal E}=0$ for $d S_\eta/d \eta$, we can write Eq.\ (\ref{I-eta}) as:
\begin{equation}
I_\eta =  \sqrt{- 2 \mu \alpha_1} \oint  \frac{d\eta }{2 \xi} \sqrt{ \frac{\alpha_3^2}{2 \mu \alpha_1} - \frac{\mu  e^2 + \alpha_2}{\mu \alpha_1} \eta - \eta^2}. 
\label{I_eta-0 loop integral}
\end{equation}
The quadratic expression in $\eta$ under the square root sign can be written as 
\begin{equation}
(\eta - \eta_{\rm min})(\eta_{\rm max} - \eta),
\label{roots eta}
\end{equation}
where $\eta_{\rm min}$ and $\eta_{\rm max}$ are the roots of the quadratic equation that we obtain by setting the expression under the square root sign in Eq.\ (\ref{I_eta-0 loop integral}) equal to zero. Eq.\ (\ref{I_eta-0 loop integral}) can thus be rewritten as
\begin{equation}
I_\eta = \sqrt{- 2 \mu \alpha_1} \int_{\eta_{\rm min}}^{\eta_{\rm max}} \frac{d\eta }{\eta} \sqrt{(\eta -\eta_{\rm min})(\eta_{\rm max} - \eta)}.
\label{I_eta-0 line integral}
\end{equation}
The quantities $(\xi_{\rm min}, \xi_{\rm max})$ in Eqs.\ (\ref{roots xi})--(\ref{I_xi-0 line integral}) and $(\eta_{\rm min}, \eta_{\rm max})$ in Eqs.\ (\ref{roots eta})--(\ref{I_eta-0 line integral}) are called the {\it turning points} of the motion or the {\it apsidal distances} \citep[p.\ 78]{Goldstein}.

To evaluate the integral in Eq.\ (\ref{I_eta-0 line integral}), we simply note that $I_\eta$ only differs from $I_\xi$ in the sign of the term with $\alpha_2$ (compare Eqs.\ (\ref{I_xi-0 loop integral}) and (\ref{I_eta-0 loop integral}) or Eqs.\ (\ref{HJ xi}) and (\ref{HJ eta}) for ${\mathcal E}=0$). In analogy with Eq.\ (\ref{I_xi-0}), we thus find
\begin{equation}
I_\eta =  \frac{\pi}{2}  \sqrt{- 2 \mu \alpha_1} \left[ - \frac{\alpha_2 + \mu e^2}{\mu \alpha_1} -\frac{2 \alpha_3}{\sqrt{- 2 \mu \alpha_1}} \right].
\label{I_eta-0}
\end{equation}

Finally, inserting $d S_\varphi/d \varphi = \alpha_3$ into Eq.\ (\ref{I-phi}), we find that
\begin{equation}
I_\varphi = 2\pi \alpha_3 = n_\varphi h.
\label{I_phi-0}
\end{equation}
In other words, $\alpha_3 = n_\varphi \hbar$ (with $\hbar \equiv h/2\pi$). In fact, $n_\varphi$ is just the absolute value of the familiar azimuthal quantum number $m$, $I_\varphi = |m|h$ and $\alpha_3 = |m| \hbar$.
  
To find $\alpha_1$ in terms of $I_\xi$, $I_\eta$, and $I_\varphi$ for the case that ${\mathcal E}=0$, we add Eqs.\ (\ref{I_xi-0}), (\ref{I_eta-0}), and (\ref{I_phi-0}). The terms with $\alpha_2$ and $\alpha_3$ cancel and the sum reduces to
\begin{equation}
I_\xi + I_\eta + I_\varphi = \pi \sqrt{- 2 \mu \alpha_1} \left( \frac{2 \mu  e^2}{-2 \mu \alpha_1} \right) =
\frac{2 \pi \mu  e^2}{\sqrt{- 2 \mu \alpha_1}}.
\label{sum I-xi I-eta I-phi 0}
\end{equation}
Solving for $\alpha_1$, we find
\begin{equation}
\alpha_1 = E = - \frac{2 \pi^2 \mu ^2 e^4}{(I_\xi + I_\eta + I_\varphi)^2}.
\label{Balmer}
\end{equation}
This is the familiar Balmer formula, as long as the principal quantum number $n$ is set equal to the sum of the quantum numbers  introduced in Eqs.\ (\ref{I-xi})--(\ref{I-phi}):
\begin{equation}
n = n_\xi + n_\eta + n_\varphi.
\label{n in old theory}
\end{equation}

To find $\alpha_2$ in terms of $I_\xi$, $I_\eta$, and $I_\varphi$, we subtract Eq.\ (\ref{I_eta-0})  from Eq.\ (\ref{I_xi-0}). The terms with $\alpha_3$ cancel and the difference reduces to
\begin{equation}
I_\xi - I_\eta = \pi \sqrt{- 2 \mu \alpha_1} \left( \frac{- 2 \alpha_2}{- 2 \mu \alpha_1} \right)
= - \frac{2 \pi \alpha_2}{\sqrt{- 2 \mu \alpha_1}}.
\label{difference I-xi I-eta 0}
\end{equation}
Using Eq.\ (\ref{sum I-xi I-eta I-phi 0}) to eliminate $\alpha_1$, we arrive at:
\begin{equation}
\alpha_2 = - \frac{\sqrt{- 2 \mu \alpha_1}}{2 \pi}  \left( I_\xi - I_\eta \right) 
=  \mu  e^2 \frac{I_\eta - I_\xi}{I_\xi + I_\eta + I_\varphi}.
\label{alpha2}
\end{equation}

We now compute the integrals for $I_\xi$, $I_\eta$, and $I_\varphi$ in Eqs.\ (\ref{I-xi})--(\ref{I-phi}) for the case that ${\mathcal E} \neq 0$.  As already noted above, the action variables $I_\xi$, $I_\eta$, and $I_\varphi$ are adiabatic invariants so their values for  ${\mathcal E} \neq 0$ are the same as for ${\mathcal E} =0$. What changes is the dependence of the separation constants $\alpha_1$, $\alpha_2$, and $\alpha_3$ on these action variables and thus the values of these separation constants, including most importantly the energy $\alpha_1$.

Since $S_\varphi$ does not depend on ${\mathcal E}$, the integral in Eq.\ (\ref{I-phi}) gives the exact same result as before: $I_\varphi = 2\pi \alpha_3 = n_\varphi h$ or $\alpha_3 = n_\varphi \hbar$ (see Eq.\ (\ref{I_phi-0})). In the cases of $I_\xi$ and $I_\eta$ it suffices to compute the integrals in Eqs.\ (\ref{I-xi})--(\ref{I-eta}) to first order in ${\mathcal E}$ as the external field will be much smaller than the Coulomb field of the nucleus. We start with $I_\xi$:
\begin{equation}
I_\xi =  \oint \frac{d S_\xi}{d \xi} \, d\xi =
\oint   \left. \left( \! \frac{d S_\xi}{d \xi} \! \right) \right|_{{\mathcal E}=0} d\xi
+   \oint \left. \frac{d}{d{\mathcal E}} \! \left( \! \frac{d S_\xi}{d \xi} \! \right)\! \right|_{{\mathcal E}=0} \! {\mathcal E}  \, d\xi
+ O({\mathcal E}^2).
\label{I-xi with E}
\end{equation}
For the first integral, we can use Eq.\ (\ref{I_xi-0}), which gives $I_\xi$ as a function of $\alpha_1$, $\alpha_2$, and $\alpha_3$ for the case that ${\mathcal E}=0$. Using Eq.\ (\ref{HJ eq xi}) for $d S_\xi/d \xi$, we can write the second integral as:
\begin{equation}
 \oint \left. \frac{d}{d{\mathcal E}} \! \left( \! \frac{d S_\xi}{d \xi} \! \right)\! \right|_{{\mathcal E}=0} \! {\mathcal E}  \, d\xi =
\oint \frac{1}{2 \xi} \frac{- \mu e \xi^3 {\mathcal E} \, d\xi}{2 \sqrt{- \alpha_3^2 + 2 (\mu  e^2 - \alpha_2) \xi + 2 \mu \alpha_1 \xi^2}}.
\label{Delta I_xi integral 1}
\end{equation}
The expression under the square root in the denominator in Eq.\ (\ref{Delta I_xi integral 1}) has the same form as the one under the square root in the numerator in Eq.\ (\ref{I_xi-0 loop integral}). It can thus be written as
\begin{equation}
\sqrt{- 2 \mu \alpha_1} \sqrt{\frac{\alpha_3^2}{2 \mu \alpha_1} - \frac{\mu  e^2 - \alpha_2}{\mu \alpha_1} - \xi^2}
= \sqrt{- 2 \mu \alpha_1}  \sqrt{(\xi -\xi_{\rm min})(\xi_{\rm max} - \xi)},
\end{equation}
with the same identification of  $\xi_{\rm min} + \xi_{\rm max}$ and $\xi_{\rm min} \xi_{\rm max}$ in terms of $\alpha_1$, $\alpha_2$, and $\alpha_3$
as before (see Eq.\ (\ref{a & b})). The second integral in Eq.\ (\ref{I-xi with E}) can then be rewritten as:
\begin{equation}
 \oint \left. \frac{d}{d{\mathcal E}} \! \left( \! \frac{d S_\xi}{d \xi} \! \right)\! \right|_{{\mathcal E}=0} \! {\mathcal E}  \, d\xi  = - \frac{\mu e {\mathcal E}}{4 \sqrt{- 2 \mu \alpha_1}} \oint \frac{\xi^2 \, d\xi}{\sqrt{(\xi - \xi_{\rm min})(\xi_{\rm max} - \xi)}}.
\label{Delta I_xi integral cont'd}
\end{equation}
The loop integral is twice the standard line integral
\begin{equation}
\int_a^b \frac{x^2 \, dx}{\sqrt{(x-a)(b-x)}} = \frac{\pi}{8} (3(a + b)^2 - 4ab).
\end{equation}
Using this result along with Eq.\ (\ref{a & b}),
we can rewrite Eq.\ (\ref{Delta I_xi integral cont'd}) as: 
\begin{equation*}
 \oint \left. \frac{d}{d{\mathcal E}} \! \left( \! \frac{d S_\xi}{d \xi} \! \right)\! \right|_{{\mathcal E}=0} \! {\mathcal E}  \, d\xi =  - \frac{\mu e {\mathcal E}}{2 \sqrt{- 2 \mu \alpha_1}} 
\frac{\pi}{8} \left[ \frac{3 ( \alpha_2 - \mu e^2 )^2}{\mu^2 \alpha_1^2} + \frac{2 \alpha_3^2}{\mu \alpha_1} \right]
\end{equation*}
\begin{equation}
= - \frac{\pi e {\mathcal E}}{16 \mu \alpha_1^2 \sqrt{- 2 \mu \alpha_1}} 
\left[ 3 ( \alpha_2 - \mu e^2 )^2 + 2 \mu \alpha_1 \alpha_3^2  \right].
\label{Delta I_xi}
\end{equation}
Since this expression is of order ${\mathcal E}$, we can use Eqs.\ (\ref{I_phi-0}), (\ref{Balmer}), and (\ref{alpha2}) giving $\alpha_1$, $\alpha_2$, and $\alpha_3$ as functions of $I_\xi$, $I_\eta$, and $I_\varphi$ {\it for the case that ${\mathcal E}=0$}. Using these three equations, we find:
\begin{equation}
( \alpha_2 - \mu e^2 ) =  \mu  e^2 \left(\frac{I_\eta - I_\xi}{I_\xi + I_\eta + I_\varphi} - 1 \right) = - \mu  e^2 \left(\frac{2I_\xi + I_\varphi}{I_\xi + I_\eta + I_\varphi} \right), 
\label{from alphas to Is (1)}
\end{equation}
\begin{equation}
2 \mu  \alpha_1 \alpha_3^2 = - \frac{\mu^2  e^4 I_\varphi^2}{(I_\xi + I_\eta + I_\varphi)^2},
\label{from alphas to Is (2)}
\end{equation}
\begin{equation}
 \alpha_1^2 \sqrt{-2\mu \alpha_1} = \left( \frac{2 \pi^2 \mu  e^4}{(I_\xi + I_\eta + I_\varphi)^2} \right)^{\!\! 2} \cdot \left( \frac{2 \pi \mu  e^2}{I_\xi + I_\eta + I_\varphi}  \right) =
\frac{8 \pi^5 \mu^3 e^{10}}{(I_\xi + I_\eta + I_\varphi)^5},
\label{from alphas to Is (3)}
\end{equation}
where in the last line we used Eq.\ (\ref{sum I-xi I-eta I-phi 0}) along with Eq.\ (\ref{Balmer}). With the help of Eqs.\ (\ref{from alphas to Is (1)}) and (\ref{from alphas to Is (2)}) we can write the final expression in Eq.\ (\ref{Delta I_xi}) as:
\begin{equation*}
- \frac{(\pi e {\mathcal E}) (\mu^2  e^4)}{16 \mu \alpha_1^2 \sqrt{- 2 \mu \alpha_1}} 
 \left[  3 \frac{(2I_\xi + I_\varphi)^2}{(I_\xi + I_\eta + I_\varphi)^2} 
 - \frac{I_\varphi^2}{(I_\xi + I_\eta + I_\varphi)^2}
 \right] 
\end{equation*}
\begin{equation}
=  - \frac{\pi e^5  \mu {\mathcal E}}{16 \alpha_1^2 \sqrt{- 2 \mu \alpha_1}} 
 \left[ \frac{12 I_\xi^2 + 12 I_\xi I_\varphi + 2 I_\varphi^2}{(I_\xi + I_\eta + I_\varphi)^2} \right].
 \label{Delta I_xi 1}
\end{equation}
Using Eq.\ (\ref{from alphas to Is (3)}), we can rewrite factor multiplying the expression in square brackets as
\begin{equation}
\frac{\pi e^5  \mu {\mathcal E}}{16} \frac{(I_\xi + I_\eta + I_\varphi)^5}{8 \pi^5 \mu^3  e^{10}} =
\frac{{\mathcal E}}{128 \pi^4 \mu^2 e^5} (I_\xi + I_\eta + I_\varphi)^5.
\label{Delta I_xi 1 prime}
\end{equation}
With the help of Eqs.\ (\ref{Delta I_xi 1}) and (\ref{Delta I_xi 1 prime}), we can rewrite Eq.\ (\ref{Delta I_xi}) as
\begin{equation}
 \oint \left. \frac{d}{d{\mathcal E}} \! \left( \! \frac{d S_\xi}{d \xi} \! \right)\! \right|_{{\mathcal E}=0} \! {\mathcal E}  \, d\xi = - \frac{{\mathcal E}}{128 \pi^4 \mu^2 e^5}  \left( I_\xi + I_\eta + I_\varphi \right)^3 \left(12 I_\xi^2 + 12 I_\xi I_\varphi + 2 I_\varphi^2 \right).
 \label{Delta I_xi 2}
\end{equation}

We can evaluate the integral (\ref{I-eta}) giving the action variable $I_\eta$ in the presence of a small external electric field ${\mathcal E}$ in the same way as we calculated the integral (\ref{I-xi}) giving $I_\xi$. In analogy with Eq.\ (\ref{I-xi with E}), we have
\begin{equation}
I_\eta =  \oint \frac{d S_\eta}{d \eta} \, d\eta =
\oint   \left. \left( \! \frac{d S_\eta}{d \eta} \! \right) \right|_{{\mathcal E}=0} d\eta
+   \oint \left. \frac{d}{d{\mathcal E}} \! \left( \! \frac{d S_\eta}{d \eta} \! \right)\! \right|_{{\mathcal E}=0} \! {\mathcal E}  \, d\eta
+ O({\mathcal E}^2).
\label{I-eta with E}
\end{equation}
For the first integral, we can use Eq.\ (\ref{I_eta-0}), which gives $I_\eta$ as a function of $\alpha_1$, $\alpha_2$, and $\alpha_3$ for the case that ${\mathcal E}=0$. Inserting Eq.\ (\ref{HJ eq eta}) for $d S_\eta/d \eta$, we can write the second integral as:
\begin{equation}
 \oint \left. \frac{d}{d{\mathcal E}} \! \left( \! \frac{d S_\eta}{d \eta} \! \right)\! \right|_{{\mathcal E}=0} \! {\mathcal E}  \, d\eta 
= \oint \frac{1}{4 \eta} \frac{\mu e \eta^3 {\mathcal E} \, d\eta}{\sqrt{- \alpha_3^2 + 2 (\mu  e^2 + \alpha_2) \eta + 2 \mu \alpha_1 \eta^2}}.
\label{Delta I_eta integral 1}
\end{equation}
Except for an overall minus sign and another minus sign in the term with $\alpha_2$, this equation has the exact same structure as Eq.\ (\ref{Delta I_xi integral 1}). Proceeding along the same lines as in Eqs.\ (\ref{Delta I_xi integral 1})--(\ref{Delta I_xi 2}), we arrive at:
\begin{equation}
\oint \left. \frac{d}{d{\mathcal E}} \! \left( \! \frac{d S_\eta}{d \eta} \! \right)\! \right|_{{\mathcal E}=0} \! {\mathcal E}  \, d\eta  = \frac{{\mathcal E}}{128 \pi^4 \mu^2  e^5}  \left( I_\xi + I_\eta + I_\varphi \right)^3 \left( 12 I_\eta^2 + 12 I_\eta I_\varphi + 2  I_\varphi^2 \right).
\label{Delta I_eta 1}
\end{equation}

To find $\alpha_1$, the total energy in the presence of a small external electric field of strength ${\mathcal E}$, we calculate, as in the case without an electric field, the sum of the action variables (to first order in ${\mathcal E}$):
\begin{equation}
I_\xi + I_\eta + I_\varphi = \frac{2 \pi \mu  e^2}{\sqrt{- 2 \mu \alpha_1}} 
+ \oint \left. \frac{d}{d{\mathcal E}} \! \left( \! \frac{d S_\xi}{d \xi} \! \right)\! \right|_{{\mathcal E}=0} \! {\mathcal E}  \, d\xi 
+ \oint \left. \frac{d}{d{\mathcal E}} \! \left( \! \frac{d S_\eta}{d \eta} \! \right)\! \right|_{{\mathcal E}=0} \! {\mathcal E}  \, d\eta
+ O({\mathcal E}^2).
\label{sum I-xi I-eta I-phi}
\end{equation}
The first term on the right-hand side is just expression (\ref{sum I-xi I-eta I-phi 0}) for the sum  $I_\xi + I_\eta + I_\varphi$ in the case that ${\mathcal E}=0$. For the next two terms, we use Eqs.\ (\ref{Delta I_xi 2}) and (\ref{Delta I_eta 1}). Noting that
\begin{equation}
I_\eta^2 +  I_\eta I_\varphi - I_\xi^2 - I_\xi I_\varphi 
= (I_\eta - I_\xi)(I_\eta + I_\xi) +  (I_\eta - I_\xi) I_\varphi 
= (I_\eta - I_\xi)(I_\eta + I_\xi + I_\varphi),
 \label{numerators}
\end{equation}
we can write these two terms as
\begin{equation}
\frac{12{\mathcal E}}{128 \pi^4 \mu^2  e^5} \left( I_\eta - I_\xi \right) \left( I_\xi + I_\eta + I_\varphi \right)^4.
\label{delta I sum}
\end{equation}
Inserting this result into Eq.\  (\ref{sum I-xi I-eta I-phi}), we arrive at:
\begin{equation}
I_\xi + I_\eta + I_\varphi = \frac{2 \pi \mu  e^2}{\sqrt{- 2 \mu \alpha_1}}
+ \frac{12{\mathcal E}}{128 \pi^4 \mu^2 e^5} \left( I_\eta - I_\xi \right) \left( I_\xi + I_\eta + I_\varphi \right)^4
+ O({\mathcal E}^2).
\end{equation}
To solve this equation for the energy $\alpha_1$ it will be helpful to schematically write it as:
\begin{equation}
x \approx \frac{y}{\sqrt{- 2 \mu \alpha_1}} + {\mathcal E}z,
\label{schematic}
\end{equation}
with the abbreviations 
\begin{equation}
x \equiv I_\xi + I_\eta + I_\varphi, \quad y \equiv 2 \pi \mu  e^2, \quad z \equiv  \frac{12}{128 \pi^4 \mu^2 e^5} \left( I_\eta - I_\xi \right) \left( I_\xi + I_\eta + I_\varphi \right)^4.
\label{abbreviations}
\end{equation}
Solving Eq.\ (\ref{schematic}) for $\alpha_1$, we find:
\begin{equation}
\alpha_1 \approx - \frac{1}{2 \mu} \left( \frac{y}{x (1 - {\mathcal E} (z/x))} \right)^2.
\label{alpha_1 abbreviated 1}
\end{equation}
To first order in ${\mathcal E}$, $\alpha_1$ is given by:
\begin{equation}
\alpha_1 \approx - \frac{1}{2 \mu} \left( \frac{y}{x (1 - {\mathcal E} (z/x))} \right)^2
\approx - \frac{y^2}{2 \mu x^2} - \frac{{\mathcal E}y^2z}{\mu x^3}.
\label{alpha_1 abbreviated 2}
\end{equation}
Inserting the expressions for $x$, $y$, and $z$ in Eq.\ (\ref{abbreviations}) into the final expression in Eq.\ (\ref{alpha_1 abbreviated 2}), we arrive at:
\begin{eqnarray}
\alpha_1 & = &  - \frac{(2 \pi \mu  e^2)^2}{2 \mu (I_\xi + I_\eta + I_\varphi)^2} - 
\left( \frac{{\mathcal E} (2 \pi \mu  e^2)^2}{\mu} \right) \left( \frac{12}{128 \pi^4 \mu^2 e^5} \right) \left( I_\eta - I_\xi \right) \left( I_\xi + I_\eta + I_\varphi \right) \nonumber \\
 & = &  - \frac{ 2 \pi^2 \mu  e^4  }{(I_\xi + I_\eta + I_\varphi)^2} +
\frac{3{\mathcal E}}{8\pi^2  e \mu} \left( I_\xi - I_\eta \right) \left( I_\xi + I_\eta + I_\varphi \right).
\label{alpha_1 final}
\end{eqnarray}
The first term on the right-hand side is just the Balmer term (cf.\ Eq.\ (\ref{Balmer})), the second term gives the first-order Stark effect, the leading term in the splitting of the energy levels by an external electric field in the $z$-direction.
Eq.\ (\ref{alpha_1 final}) is exactly the result reported by \citet[p.\ 508, Eq.\ (62)]{Epstein 1916a} and by \citet[p.\ 18, Eq.\ (46)]{Kramers 1919}. Imposing the quantum conditions (\ref{I-xi})--(\ref{I-phi}) and setting the sum $I_\xi + I_\eta + I_\varphi$ equal to the principal quantum number $n$ times $h$, we find that the energy levels $E = \alpha_1$ of a hydrogen atom in an electric field are given by
\begin{equation}
E = - \frac{\mu  e^4}{2 n^2 \hbar^2} + \frac{3{\mathcal E}\hbar^2}{2   e \mu} \left( n_\xi - n_\eta \right)n.
\label{Stark formula}
\end{equation}
The splitting of the energy levels of the first-order Stark effect in hydrogen is thus proportional to ${\mathcal E} (n_\xi - n_\eta)n$.\footnote{See Fig.\ 1 in \citet{Duncan and Janssen 2014} for a graphical representation, based on tables given by \citet[pp.\ 512--513]{Epstein 1916b}, of the splittings both of the energy levels with principal quantum number $n=2$ and $n=3$ and of the Balmer line $H_\alpha$ associated with transitions between those energy levels.} Most experts agreed that the experimental data on the Stark effect were in excellent agreement with this formula, even though Stark, a staunch opponent of the old quantum theory, did not \citep[pp.\ 127--128, 168--169]{Kragh 2012}.

\begin{figure}[h]
   \centering
   \includegraphics[width=4in]{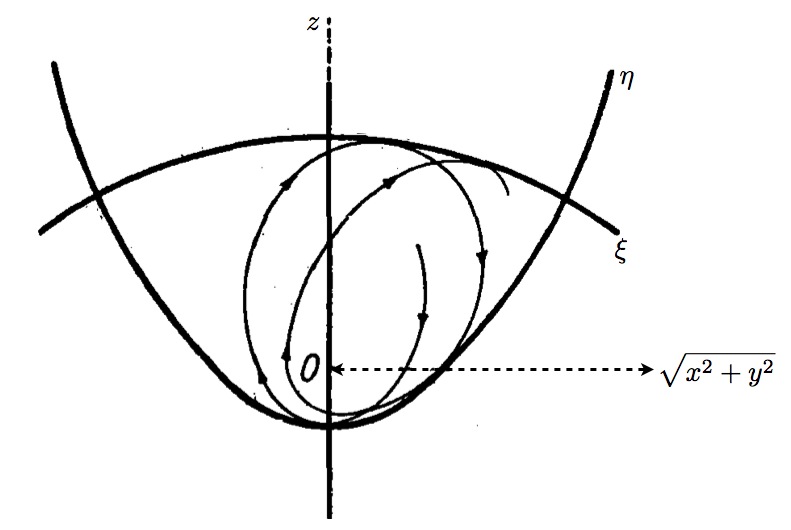} 
   \caption{{\small The problem with orbits with $n_\varphi = 0$. This figure is taken from \citet[p.\ 499]{Epstein 1916b} but relabeled the same way as Fig.\ \ref{parabolic}.}}
   \label{pendelbahn}
\end{figure}

As mentioned in the introduction, we focus on two problems with Epstein and Schwarz\-schild's account of the Stark effect on the basis of the Bohr-Sommerfeld theory. First, as we will show in section 3, the orbits found in parabolic coordinates do not reduce to the orbits found in spherical coordinates if we let the strength ${\mathcal E}$ of the external field go to zero. Second, this account of the Stark effect calls for extra restrictions on the allowed orbits over and above the quantum conditions (\ref{I-xi})--(\ref{I-phi}). In wave mechanics, by contrast, the basic quantum condition (i.e., the normalizability of bound-state eigenfunctions) suffices to get the correct formula for the Stark effect. In section 4, in which we will cover  Schr\"odinger and Epstein's account of the Stark effect on the basis of wave mechanics, we will encounter the same quantum numbers $n_\xi$, $n_\eta$, and $n_\varphi$ that we encountered above (see Eqs.\ (\ref{I-xi})--(\ref{I-phi})). However, the relation between these three quantum numbers and the principal quantum number $n$ is different in the two theories. In the old quantum theory, as we saw above (see Eqs.\ (\ref{alpha_1 final})--(\ref{Stark formula})), the relation is given by:
\begin{equation}
n = n_\xi + n_\eta + n_\varphi.
\label{n in old quantum theory}
\end{equation}
In wave mechanics it gets replaced by
\begin{equation}
n = n_\xi + n_\eta + n_\varphi + 1.
\label{n in wave mechanics}
\end{equation}
The additional term of 1 obviates the need for extra restrictions required by the old quantum theory. In section 5, we use the WKB method of finding approximate solutions of the Schr\"odinger equation to derive amended versions of the quantum conditions in the old quantum theory, which changes Eq.\ (\ref{n in old quantum theory}) to Eq.\ (\ref{n in wave mechanics}) so that extra conditions are no longer needed.

To conclude this section, we review what these extra restrictions are. There are two of them. First, it immediately follows from Eq.\ (\ref{n in old quantum theory}) that the three quantum numbers  $n_\xi$, $n_\eta$, and $n_\varphi$ cannot simultaneously be zero, as that would result in the energy, $E \propto - 1/n^2$, becoming infinite.  Second, even in cases where $n_\xi$ and/or $n_\eta$ are non-zero, so that the energy is finite, a zero value for $n_\varphi$ still needs to be ruled out as that value would give rise to an unstable orbit. The problem is illustrated in Fig.\ \ref{pendelbahn}. If $n_\varphi = 0$, the orbit of the electron would be in the plane in the figure. Contrary to the situation depicted in Fig.\ \ref{parabolic}, this plane would not be rotating around the $z$-axis. As long as there is no external electric field, such orbits are allowed. However, as soon as the electric field in the $z$-direction is switched on, such orbits become unstable and the electron will eventually hit the nucleus. 

\section{The non-uniqueness of orbits in the old quantum theory}

In this section, we compare the orbits found by applying the quantum conditions of the old quantum theory to the motion of the electron in a hydrogen atom without an external electric field in two different sets of coordinates, spherical coordinates and parabolic coordinates. In the presence of an external field, we cannot use spherical coordinates because the Hamilton-Jacobi equation for the system is not separable in those coordinates. This is why, in section 2, we used parabolic coordinates instead. As long as there is no electric field, however, the Hamilton-Jacobi equation is separable in both sets of coordinates. Proceeding in the same way that got us to Eqs.\ (\ref{I-xi})--(\ref{I-phi}) in section 2, we set the action variables corresponding to the spherical coordinates $(r, \vartheta, \varphi)$ equal to integral multiples of $h$ \citep[sec.\ 64, pp.\ 200--206]{Corben and Stehle}:
\begin{eqnarray}
I_r = \oint p_r \, dr = n_r h \label{I-spherical-r},  \\
I_\vartheta = \oint p_\vartheta \, d\vartheta = n_\vartheta h \label{I-spherical-theta}, \\
I_\varphi = \oint p_\varphi \, d\varphi = n_\varphi h \label{I-spherical-phi},
\end{eqnarray}
where, at least at this point, $n_r$, $n_\vartheta$, and $n_\varphi$ can take on the values $0, 1, 2, \ldots$. Examining the relation between  quantum numbers and  separation constants in this case (cf.\ Eqs.\ (\ref{I-spherical-phi})--(\ref{n in old theory}) in section 2), we find the relation between the quantum numbers $(n_r, n_\vartheta, n_\varphi)$ and the usual quantum numbers $(n, l, m)$. As in the case of parabolic coordinates (see Eq.\ (\ref{Balmer})), the energy $E$ depends on the sum of the action variables:
\begin{equation}
E = -\frac{2 \pi^2 \mu e^4}{(I_r + I_\vartheta + I_\varphi)^2} = -\frac{2 \pi^2 \mu e^4}{(n_r + n_\vartheta + n_\varphi)^2 h^2}
\label{Balmer spherical}
\end{equation}
It follows that the principal quantum number $n$ is given by
\begin{equation}
n = n_r + n_\vartheta + n_\varphi,
\label{principal spherical}
\end{equation}
and that $n_r$, $n_\vartheta$, and $n_\varphi$ cannot all be zero simultaneously.
The  total angular momentum, $L = l \hbar$, and the $z$-component of the angular momentum, $L_z = m \hbar$, it turns out, are equal to $(1/2\pi)$ times $I_\vartheta + I_\varphi$ and $I_\varphi$, respectively. It follows that 
\begin{equation}
l = n_\vartheta + n_\varphi, \quad  n_\varphi = |m|.
\label{l and m}
\end{equation}
The total angular momentum cannot be zero as that would correspond to straight-line motion through the nucleus. Like the principal quantum number $n$, the quantum number $l$ can thus only take on the value $1, 2, 3, \ldots$. The quantum number $m$ can take on the values $-l, \ldots 0, \ldots l$. Comparing Eqs.\ (\ref{principal spherical}) and (\ref{l and m}), we note that
\begin{equation}
n = n_r + l.
\label{n_r + l}
\end{equation}
The orbits are ellipses characterized by their semi-major axes $a$ and their eccentricities $\epsilon$. In analogy to the minimum and maximum distance of a planet to the Sun at perihelion and aphelion, respectively, we consider the {\it apsidal distances}  $r_{\rm max}$ and $r_{\rm min}$, the maximum and minimum distances of the electron to the nucleus at one of the focal points of the ellipse. These apsidal distances can be expressed in terms of $a$ and $\epsilon$ \citep[p.\ 96]{Goldstein}:
\begin{equation}
 r_{\rm max} = a(1+\epsilon), \quad r_{\rm min} =  a(1-\epsilon).
\end{equation}
Inverting these relations, we find
\begin{eqnarray}
a & = & {\textstyle \frac{1}{2}} (r_{\rm max}+r_{\rm min}), \label{a} \\
 & & \nonumber \\
\epsilon & = & \frac{r_{\rm max}-r_{\rm min}}{r_{\rm max}+r_{\rm min}}. \label{epsilon}
\end{eqnarray}
The eccentricity can also be expressed in terms of the energy $E$ and the angular momentum $L$ (cf.\ Goldstein p.\ 94, Eq.\ (3.57); Corben and Stehle, 1994, p.\ 95, Eq.\ (37.8)):
\begin{equation}
\epsilon = \sqrt{1 + \frac{2EL}{\mu e^4}}.
\label{epsilon in E and L}
\end{equation}
Inserting Eq.\ (\ref{Balmer spherical}) for $E$ and $l\hbar$ for $L$, we find that this equation reduces to:
\begin{equation}
\epsilon = \sqrt{1 - \frac{l^2}{n^2}}.
\label{epsilon in n and l}
\end{equation}
This equation also allows us to compute $l$ once we know $n$ and $\epsilon$:
\begin{equation}
 l = n\sqrt{1-\epsilon^{2}}.
\label{l in epsilon and n}
\end{equation}

Table 1 shows the values of the angular momentum $L$ (in terms of $l = L \hbar$) and the eccentricity ($\epsilon$) for orbits corresponding to low values for the principal quantum number $n = n_r + l$, if we select the allowed orbits by imposing the quantum conditions in spherical coordinates.
  
 \begin{table}[h]
\centering
\label{Table 1}
\begin{tabular}{|c|c|c|c|c|c|}
\hline
\multicolumn{1}{|c|}{$n$}
&\multicolumn{1}{c|}{ $n_{r}$}
&\multicolumn{1}{c|}{ $l$}
&\multicolumn{1}{c|}{ $\epsilon$} \\[2pt]
\hline
1  & 0  & 1 & 0   \\
2  & 0 & 2 & 0  \\
2  & 1 & 1 & $\sqrt{3}/2$ \\
3  & 0 & 3 & 0  \\
3  & 1 & 2 & $\sqrt{5}/3$ \\
3  & 2 & 1 & $2\sqrt{2}/3$  \\
\hline
\end{tabular}
\vspace{.1in}
\caption{{\small Parameters characterizing orbits in hydrogen for low values of the principal quantum number $n = n_r + l$ (see Eq.\ (\ref{n_r + l})) when quantum conditions are imposed in {\bf spherical coordinates}: angular momenta $L$ (in terms of $l = L/ \hbar$) and eccentricities $\epsilon$ (see Eq.\ (\ref{epsilon in n and l})).}}
\end{table}

We now compare the angular momenta and eccentricities of these orbits with the angular momenta and eccentricities of orbits that we find if we impose the quantum conditions in parabolic coordinates $(\xi, \eta, \varphi)$ (see Eq.\ (\ref{parabolic coordinates})). The results are collected in Table 2. 

We will only present the calculations for the special case that one of the two quantum numbers $n_\xi$ or $n_\eta$ equals zero. Suppose $n_\eta = 0$. The case that $n_\xi =0$ can be handled in the exact same way. If $n_\eta = 0$, the action variable $I_\eta = n_\eta h = 0$. This action variable is essentially equal to the integral of an expression of the form $\sqrt{(\eta -\eta_{\rm min})(\eta_{\rm max} - \eta)}$ from $\eta_{\rm min}$ to $\eta_{\rm max}$ (see Eq.\ (\ref{I_eta-0 line integral})). This integral  vanishes only if the integrand is identically zero, i.e., if the value of $\eta$ is fixed: $\eta = \eta_{\rm min} = \eta_{\rm max}$. In section 2, we saw that $\eta_{\rm min}$ and  $\eta_{\rm max}$ are the roots of the quadratic equation obtained by setting the expression under the square root sign in the integral giving the action variable $I_\eta$ equal to zero:
\begin{equation}
\frac{\alpha_3^2}{2 \mu \alpha_1} - \left( \frac{\mu  e^2 + \alpha_2}{\mu \alpha_1} \right) \eta - \eta^2 = 0
\end{equation}
(cf.\ Eq.\ (\ref{I_eta-0 loop integral})). If the two roots of this equation, which is of the form $a \eta^2 + b \eta + c =0$, are the same, the discriminant, $b^2 -4ac$, vanishes and the roots are simply equal to $-b/2a$. In other words,
\begin{equation}
\eta_{\rm min} = \eta_{\rm max} = - \frac{\mu  e^2 + \alpha_2}{2\mu \alpha_1}.
\label{eta fixed}
\end{equation}
With the help of Eqs.\ (\ref{Balmer}) and (\ref{alpha2}), $\alpha_1$ and $\alpha_2$ can be expressed in terms of the action variables $(I_\xi, I_\eta, I_\varphi)$ and the quantum numbers $(n_\xi, n_\eta, n_\varphi = |m|)$ associated with them.

If $\eta$ is constant, then $r = (\xi + \eta)/2$ (see Eq.\ (\ref{r})) reaches its minimum and maximum value  whenever $\xi$ reaches its minimum and maximum value. In section 2, we saw that $\xi_{\rm min}$ and $\xi_{\rm max}$ are the roots of the quadratic equation obtained by setting the expression under the square root sign in the integral giving the action variable $I_\xi$ equal to zero:
\begin{equation}
\frac{\alpha_3^2}{2 \mu \alpha_1} - \left( \frac{\mu  e^2 - \alpha_2}{\mu \alpha_1} \right) \xi - \xi^2 = 0
\end{equation}
(cf.\ Eq.\ (\ref{I_xi-0 loop integral})). Solving this equation gives us expressions for $\xi_{\rm min}$ and $\xi_{\rm max}$ in terms of the separation constants $(\alpha_1, \alpha_2,  \alpha_3)$. We can then use Eqs.\ (\ref{Balmer}), (\ref{alpha2}), and (\ref{I_phi-0}) to express $(\alpha_1, \alpha_2,  \alpha_3)$ in terms of the quantum numbers $(n_\xi, n_\eta, n_\varphi = |m|)$.  Finally, we set $n_\eta = 0$ and derive expressions for $r_{\rm max} =  \frac{1}{2} (\xi_{\rm max}  +  \eta_{\rm max})$ and $r_{\rm min} = \frac{1}{2} (\xi_{\rm min} + \eta_{\rm min})$, using Eq.\ (\ref{eta fixed}) for $\eta_{\rm min} = \eta_{\rm max}$. In this way we find:
\begin{eqnarray}
r_{\rm min} & = & \frac{n h^2}{\mu \pi^2 e^2} \left( n - \sqrt{n_\xi (n_\xi + |m|) } \right),  \nonumber \\
 & & \label{xi-min and xi-max} \\
r_{\rm max} & = & \frac{n h^2}{\mu \pi^2 e^2} \left( n + \sqrt{n_\xi (n_\xi + |m|) } \right). \nonumber
\end{eqnarray}
Inserting these results into Eq.\ (\ref{epsilon}) for $\epsilon$, we find:
\begin{equation}
\epsilon = \frac{r_{\rm max}-r_{\rm min}}{r_{\rm max}+r_{\rm min}} = \sqrt{\frac{n_\xi (n_\xi + |m|) }{n^2}}.
\label{epsilon parabolic eta = 0}
\end{equation}
A similar results holds for orbits for which $n_\xi =0$:
\begin{equation}
\epsilon = \sqrt{\frac{n_\eta (n_\eta + |m|) }{n^2}}.
\label{epsilon parabolic xi = 0}
\end{equation}

If both $n_\xi$ and $n_\eta$ are non-zero, it is more difficult to determine $r_{\rm min}$ and $r_{\rm max}$. If $\xi$ oscillates between $\xi_{\rm min}$ and $\xi_{\rm max}$ and $\eta$ oscillates between $\eta_{\rm min}$ and $\eta_{\rm max}$, it depends on the phase difference $\delta$ between those two oscillations when $r = (\xi + \eta)/2$ reaches its minimum and maximum value and what those values are. We will not present the calculations of the eccentricities for such orbits but simply state the result. For arbitrary values of $n_\xi$ and $n_\eta$, the eccentricity is given by
\begin{equation}
\epsilon = \sqrt{\sigma_1^2+\sigma_2^2+2\sigma_1 \sigma_2 \cos{(2\pi\delta)}},
\label{epsilon parabolic nxi and neta neq 0}
\end{equation}
where $\sigma_1$ and $\sigma_2$ are defined as:
\begin{equation}
\sigma_1 \equiv \frac{1}{n}\sqrt{n_\xi(n_\xi+ |m|)}, \quad  \sigma_2 \equiv \frac{1}{n}\sqrt{n_\eta(n_\eta+ |m|)}.
\label{sigma}
\end{equation}
For $n_\eta =0$ and $n_\xi =0$, Eq.\ (\ref{epsilon parabolic nxi and neta neq 0}) reduces to Eq.\ (\ref{epsilon parabolic eta = 0}) and Eq.\ (\ref{epsilon parabolic xi = 0}), respectively.

Table 2 brings together the results for all possible values of $n_\xi$ and $n_\eta$ for orbits corresponding to low values of $n = n_\xi + n_\eta + |m|$ (cf.\ Eq.\ (\ref{n in old theory}) with $n_\varphi = |m|$), if we select the allowed orbits by imposing the quantum conditions in parabolic coordinates. The table shows the values of the eccentricity $\epsilon$ and the angular momentum $L$ (in terms of $l = L \hbar$) for these orbits. For those orbits for which either  $n_\eta =0$ or $n_\xi = 0$, we used Eqs.\ (\ref{epsilon parabolic eta = 0}) and (\ref{epsilon parabolic xi = 0}), respectively, to compute $\epsilon$. In all cases, we inserted the values for $\epsilon$ into Eq.\ (\ref{l in epsilon and n}) to compute $l$.  

\begin{table}[h]
\centering
\label{Table 2}
\begin{tabular}{|c|c|c|c|c|c|}
\hline
\multicolumn{1}{|c|}{$n$}
&\multicolumn{1}{|c|}{$n_\xi$}
&\multicolumn{1}{|c|}{$n_\eta$}
&\multicolumn{1}{c|}{ $|m|$}
&\multicolumn{1}{c|}{ $l$}
&\multicolumn{1}{c|}{ $\epsilon$}
 \\[2pt]
\hline
1  & 0  & 0 & 1 & 1 & 0   \\
2  & 0 & 0  & 2 & 2 & 0 \\
2  & 1 & 0 & 1 & $\sqrt{2}$ & $1/\sqrt{2}$   \\
2  &  0 & 1 & 1& $\sqrt{2}$ & $1/\sqrt{2}$  \\
2 & 1 & 1 & 0 &  $2\sin{(\pi\delta)}$ & $\cos{(\pi\delta)}$  \\
3  & 0 & 0 & 3 & 3 & 0   \\
3  & 1 & 1 & 1 & $\sqrt{1+8\sin^{2}{(\pi\delta)}}$ & $2\sqrt{2}\cos{(\pi\delta)}/3$   \\
3  & 2 & 0 & 1 & $\sqrt{3}$ & $\sqrt{2/3}$   \\
3  & 0 & 2 & 1 & $\sqrt{3}$ & $\sqrt{2/3}$ \\
3  & 1 & 0 & 2 & $\sqrt{6}$ & $1/\sqrt{3}$   \\
3  & 0 & 1 & 2 & $\sqrt{6}$ & $1/\sqrt{3}$  \\
\hline
\end{tabular}
\vspace{.1in}
\caption{{\small Parameters characterizing orbits in hydrogen for low values of the principal quantum number $n = n_\xi + n_\eta + |m|$ (cf.\ Eq.\ (\ref{n in old theory}) with $n_\varphi = |m|$) when quantum conditions are imposed in {\bf parabolic coordinates}: angular momenta $L$ (in terms of  $l = L/\hbar$) and eccentricities $\epsilon$ (computed on the basis of Eqs.\ (\ref{epsilon parabolic eta = 0})--(\ref{sigma})).}}
\end{table}

Comparison of Tables 1 and 2 clearly shows that the allowed orbits depend on the coordinates in which the quantum conditions are imposed. For circular orbits, for which the eccentricity $\epsilon$ equals zero, the orbits are the same. In all other cases, the orbits are different. If the quantum conditions are imposed in spherical coordinates, the angular momentum is always an integral multiple of $\hbar$. If they are imposed in parabolic coordinates, this is true only for circular orbits. The angular momentum is not even discrete in all cases. If both $n_\xi$ and $n_\eta$ are non-zero, it can take on a continuous range of values labeled by the phase factor $\delta$.\footnote{If the quantum conditions are imposed in so-called prolate spheroidal coordinates, another coordinate system in which the Hamilton-Jacobi equation for an electron in a hydrogen atom without external electric field is also separable, we find yet another set of different orbits. We leave this as an exercise to the reader.}

Both \citet[p.\ 507]{Epstein 1916b} and \citet[p.\ 284]{Sommerfeld 1923} acknowledged  the problem with the non-uniqueness of the orbits.\footnote{See \citet{Duncan and Janssen 2014} for the passages in which Epstein and Sommerfeld stated the problem. The problem is also noted by \citet[p.\ 121]{Juvet 1926}, professor at the University of Neuchatel, in a book on analytical mechanics and the old quantum theory.} The solution they proposed was little more than wishful thinking. Their hope was that, if relativistic effects were included in the Hamiltonian in Eq.\ (\ref{Hamiltonian0}), there would only be one coordinate system left in which the Hamilton-Jacobi equation would be separable. The actual orbits would then be the orbits found if the quantum conditions were imposed in those coordinates. Unfortunately, as \citet{Einstein 1917} suspected early on, there is in general simply no coordinate system in which the Hamilton-Jacobi equation for a Hamiltonian containing a variety of physical effects is separable. So the solution suggested by Epstein and Sommerfeld does not work. There are either two or more coordinate systems in which the Hamilton-Jacobi can be separated or none at all.


\section{The Stark effect in wave mechanics}

Shortly after \citet{Schroedinger 1926a} introduced wave mechanics,  \citet{Schroedinger 1926b} and \citet{Epstein 1926},  independently of one another, applied the new wave mechanics to the Stark effect.\footnote{Schr\"odinger's paper was received by {\it Annalen der Physik} on May 10 and published July 13, 1926. Epstein's paper is signed July 29 and appeared in {\it Physical Review} in October 1926. Epstein had moved from Munich to Pasadena in 1921. In his paper, \citet[p.\ 695, note 1]{Epstein 1926} cited Schr\"odinger's first and second ``communication'' ({\it Mitteilung}) on wave mechanics but not the third. Presumably, the July 13 issue of {\it Annalen der Physik} had not reached Pasadena by July 29.} For our purposes, it suffices to solve the Schr\"odinger equation for the hydrogen atom {\it without} an external electric field in parabolic coordinates. This amounts to a derivation of the extra term of 1 in the relation (\ref{n in wave mechanics}) between the principal quantum number and the quantum numbers in parabolic coordinates that we mentioned at the end of section 2. We will only sketch the derivation of the actual formula for the first-order Stark effect, i.e., the formula for the energy levels in the presence of an external electric field. 

As in the old quantum theory, the starting point for the derivation of the formula for the Stark effect in hydrogen is the Hamiltonian (\ref{Hamiltonian}) in parabolic coordinates.\footnote{For the main steps in the derivation below, see \citet[pp.\ 398--399]{Condon and Shortley 1963}.} Instead of the substitutions (\ref{p to dS/dq}) of $\partial S/\partial \xi$ for $p_\xi$ etc., we now make the substitutions
\begin{equation}
p_\xi \longrightarrow \frac{\hbar}{i} \frac{\partial}{\partial \xi}, \;\;\; p_\eta \longrightarrow \frac{\hbar}{i} \frac{\partial}{\partial \eta} \;\;\; p_\varphi \longrightarrow \frac{\hbar}{i} \frac{\partial}{\partial \varphi},
\label{p to operator p}
\end{equation}
to form the Hamilton operator entering into the time-independent Schr\"odinger equation
\begin{equation}
H \psi = \alpha_1 \psi,
\label{Schroedinger eq}
\end{equation}
where $\psi(\xi, \eta, \varphi)$ is the wave function in parabolic coordinates and where, to bring out the family resemblance of the calculation in wave mechanics to the one in the old quantum theory given in section 2, we used $\alpha_1$ to label the eigenvalues of energy. With the substitutions (\ref{p to operator p}) the Hamiltonian (\ref{Hamiltonian}) becomes:
\begin{equation}
H = - \frac{\hbar^2}{2\mu} \left( \frac{4}{\xi + \eta} \left( \frac{\partial}{\partial \xi} \xi \frac{\partial}{\partial \xi} \right) +  \frac{4}{\xi + \eta} \left( \frac{\partial}{\partial \eta} \eta \frac{\partial}{\partial \eta} \right) + \frac{1}{\xi \eta} \frac{\partial^2}{\partial \varphi^2}  \right)
- \frac{2e^2}{\xi + \eta} +  \frac{1}{2} e{\mathcal E}(\xi - \eta).
\label{Hamilton operator}
\end{equation}
Inserting this Hamilton operator into Eq.\ (\ref{Schroedinger eq}), dividing both sides by $\psi$ and multiplying by $2\mu (\xi + \eta)$ (using relations (\ref{xi & eta})), we arrive at the Schr\"odinger equation:
\begin{equation}
- \frac{\hbar^2}{\psi} \left( 4 \frac{\partial}{\partial \xi} \xi \frac{\partial}{\partial \xi} +  4   \frac{\partial}{\partial \eta} \eta \frac{\partial}{\partial \eta} +   \left( \frac{1}{\xi} + \frac{1}{\eta}  \right) \frac{\partial^2}{\partial \varphi^2} \right) \psi - 4 \mu e^2  +  \mu e {\mathcal E}(\xi^2 - \eta^2)  =  2\mu (\xi + \eta) \alpha_1.
\label{Schroedinger eq H hydrogen}
\end{equation}
Note the similarity between this equation and the Hamilton-Jacobi equation (\ref{HJ eq}). The Schr\"odinger equation, like the Hamilton-Jacobi equation for this system, is separable in parabolic coordinates. 
In the case of the Schr\"odinger equation, this means that its solution factorizes as follows:\footnote{Condon and Shortley use $F$ and $G$ for what we call $\psi_\xi$ and $\psi_\eta$, respectively.}
\begin{equation}
\psi(\xi, \eta, \varphi) = \psi_\xi(\xi) \psi_\eta(\eta) \psi_\varphi(\varphi).
\label{factorization}
\end{equation}
The wave function $\psi$ and the generating function $S$ are related via $\psi = e^{{\displaystyle iS/\hbar}}$. Hence, if $S$ is the {\it sum} of three functions, each of which depends on only one of the three coordinates $(\xi, \eta, \varphi)$, $\psi$ must be the {\it product} of three such functions:
\begin{equation}
\psi(\xi, \eta, \varphi) = e^{i(S_\xi(\xi) + S_\eta(\eta) + S_\varphi(\varphi))/\hbar} = \psi_\xi(\xi) \psi_\eta(\eta) \psi_\varphi(\varphi),
\label{psi & S}
\end{equation}
with 
\begin{equation}
\psi_\xi(\xi) = e^{iS_\xi(\xi)/\hbar}, \;\;\; \psi_\eta(\eta) = e^{iS_\eta(\eta)/\hbar}, \;\;\; \psi_\varphi(\varphi) = e^{iS_\varphi(\varphi)/\hbar}.
\label{factors}
\end{equation}
Just as we could set $S_\varphi(\varphi)$ equal to $\alpha_3 \varphi$, we can now set $\psi_\varphi(\varphi)$ equal to $e^{i\alpha_3 \varphi/\hbar}$, with $\alpha_3 = |m| \hbar$ (cf. Eq.\ (\ref{I_phi-0})). After we substitute $-m^2 \psi$ for $\partial^2 \psi/\partial \varphi^2$ in Eq.\ (\ref{Schroedinger eq H hydrogen}), we are left with an equation that splits into a part that only depends on $\xi$ and a part that only depends on $\eta$. The two parts must therefore each be constant. Denoting these constants by $\mp 2 \alpha_2$ as we did in the corresponding Eqs.\ (\ref{HJ xi})--(\ref{HJ eta}) in the old quantum theory, we arrive at
\begin{equation}
- \frac{4 \hbar^2}{\psi_\xi} \frac{d}{d \xi} \left( \xi \frac{d\psi_\xi}{d \xi} \right) + \frac{m^2 \hbar^2}{\xi}  - 2 \mu  e^2  +  \mu e {\mathcal E} \xi^2 -  2 \mu \alpha_1 \xi = - 2 \alpha_2,
\label{psi xi eq}
\end{equation}
\begin{equation}
- \frac{4\hbar^2}{\psi_\eta} \frac{d}{d \eta} \left( \eta \frac{d\psi_\eta}{d \eta} \right) + \frac{m^2 \hbar^2}{\eta}  - 2 \mu  e^2  -  \mu e{\mathcal E} \eta^2 - 2 \mu \alpha_1 \eta = + 2 \alpha_2.
\label{psi eta eq}
\end{equation}
These last two equations can be rewritten as
\begin{equation}
\xi  \frac{d^2\psi_\xi}{d \xi^2} + \frac{d\psi_\xi}{d \xi} + \frac{1}{4}
\left[ 
\frac{2 \mu  e^2}{\hbar^2} - \frac{2 \alpha_2}{\hbar^2} - \frac{m^2}{\xi} 
+ \frac{2 \mu \alpha_1}{\hbar^2} \xi - \frac{\mu e {\mathcal E}}{\hbar^2} \xi^2 \right] \psi_\xi = 0,
\label{psi xi eq 2}
\end{equation}
\begin{equation}
\xi  \frac{d^2\psi_\eta}{d \eta^2} + \frac{d\psi_\eta}{d \eta} + \frac{1}{4}
\left[ 
\frac{2 \mu  e^2}{\hbar^2} + \frac{2 \alpha_2}{\hbar^2} - \frac{m^2}{\eta} 
+ \frac{2 \mu \alpha_1}{\hbar^2} \eta + \frac{\mu e {\mathcal E}}{\hbar^2} \eta^2
\right] \psi_\eta =0.
\label{psi eta eq 2}
\end{equation}
We now solve Eq.\ (\ref{psi xi eq 2}) for $\psi_\xi(\xi)$ for the case that ${\mathcal E}=0$. It will be convenient to introduce the quantity $n$ defined as
\begin{equation}
na \equiv \frac{\hbar}{\sqrt{-2 \mu \alpha_1}},
\label{n & a}
\end{equation}
where $a \equiv \hbar^2/\mu e^2$ is the Bohr radius. The notation $n$ was chosen with malice aforethought as $n$ will turn out to be the principal quantum number (cf.\ Eqs.\ (\ref{sum I-xi I-eta I-phi 0})--(\ref{n in old theory})). Eq.\ (\ref{n & a}) allows us to write 
\begin{equation}
\alpha_1 = - \frac{\hbar^2}{2 \mu a^2 n^2}.
\label{alpha1 & na}
\end{equation}
With the help of Eqs.\ (\ref{n & a}) and (\ref{alpha1 & na}), Eq.\ (\ref{psi xi eq 2}) for ${\mathcal E}=0$ can be rewritten as:
\begin{equation}
\xi  \frac{d^2\psi_\xi}{d \xi^2} + \frac{d\psi_\xi}{d \xi} + \frac{1}{4}
\left[   \frac{2}{a} - \frac{2 \alpha_2}{\hbar^2}  - \frac{m^2}{\xi} - \frac{\xi}{n^2 a^2} \right] \psi_\xi = 0.
\label{psi xi eq 3}
\end{equation}
To solve this equation we first need to examine its asymptotic behavior. The upshot of that is that  $\psi_\xi$ can be written in the form
\begin{equation}
\psi_\xi(\xi) = \xi^{|m|/2} e^{- \xi/2na} f(\xi),
\label{psi xi factorized}
\end{equation}
where $f(\xi)$ should be finite as $\xi\rightarrow 0$ and power bounded for $\xi\rightarrow\infty$. Given this factorization, we can write the first derivative of $\psi_\xi$ and  $\xi$ times the second derivative of $\psi_\xi$ as:\footnote{In Eq.\ (\ref{psi xi factorized}), $\psi_\xi$ is written as the product of three factors. Using 0s, 1s, and 2s to indicate which factors are differentiated zero, one, and two times, we can schematically write $d\psi_\xi/d \xi$ as $[100 + 010 + 001] \psi_\xi$,
and $d^2\psi_\xi/d \xi^2$ as $[(200 + 110 + 101) + (110 + 020 + 011) + (101 + 011 + 002)]\psi_\xi = [200 + 2 (110) + 2 (101) + 020 + 2 (011) + 002] \psi_\xi$.}
\begin{equation}
\frac{d\psi_\xi}{d \xi} = \left( \frac{|m|}{2} \xi^{ |m|/2 - 1} f - \frac{1}{2na} \xi^{|m|/2} f +  \xi^{|m|/2} f' \right) e^{- \xi/2na},
\label{psi xi '}
\end{equation}
\begin{eqnarray}
\xi \frac{d^2\psi_\xi}{d \xi^2} & = & \left( {\textstyle \frac{|m|}{2}} \left( 
{\textstyle \frac{|m|}{2}} - 1 \right)  \xi^{|m|/2 - 1} f
-    \frac{|m|  \xi^{|m|/2}}{2na} f
+ |m| \xi^{|m|/2} f'
\right.  \nonumber \\
  &  & \left. \;\;\;\; \;\;\;\;  \;\;\;\; \;\;\;\; 
+ \;  \frac{\xi^{|m|/2 + 1}}{4n^2a^2} f
\; - \; \frac{\xi^{|m|/2 + 1}}{na} f'
\; + \;  \xi^{|m|/2 + 1} f''
  \right) e^{-\xi/2na}.
\label{psi xi ''}
\end{eqnarray}
When these results are inserted into Eq.\ (\ref{psi xi eq 3}), the last two terms in $\frac{1}{4} [\ldots] \psi_\xi$ cancel against the terms $(|m|^2/4\xi) \psi_\xi$ and $(1/4n^2a^2) \psi_\xi$ in $\xi \, d^2\psi_\xi/d \xi^2$; and the first term in $d\psi_\xi/d \xi$ cancels against the term $-(|m|/2) \psi_\xi/\xi$ in $\xi \, d^2\psi_\xi/d \xi^2$. Dividing the remainder of Eq.\ (\ref{psi xi eq 3}) by $\xi^{|m|/2} e^{-\xi/2na}$ and grouping terms with $f$, $f'$, and $f''$, we find:
\begin{equation}
\xi f'' + \left( |m| + 1 - \frac{\xi}{na}  \right) f' + 
\left( \frac{1}{2a} \left(1 - \frac{1}{n} \right) - \frac{\alpha_2}{2 \hbar^2} - \frac{|m|}{2na} \right) f = 0.
\label{eq for f}
\end{equation}
The solution of this equation will be a polynomial in $\xi$:
\begin{equation}
f(\xi) = \sum_k c_k \xi^k.
\label{polynomial}
\end{equation}
Inserting this polynomial into Eq.\ (\ref{eq for f}), we find:
\begin{equation}
\sum_k c_k \left(  k(k \! - \! 1) \xi^{k-1} \! + \! (|m| \! + \! 1) k  \xi^{k-1}  \! - \!  \frac{k }{na} \xi^k \! + \!
\left( \frac{1}{2a} \left(1 \! - \!  \frac{|m| \! + \! 1}{n} \right)  \! - \!  \frac{\alpha_2}{2 \hbar^2}  \right) \xi^k \right) = 0.
\label{polynomial into eq for f}
\end{equation}
Replacing the summation variable $k$ by $k \! + \! 1$, we can rewrite the terms of order $k \! - \! 1$ in $\xi$ in Eq.\ (\ref{polynomial into eq for f}) as terms of order $k$:
\begin{equation}
\sum_k k c_k (k  + |m|) \xi^{k-1} =  \sum_k (k+1) c_{k \! + \! 1} (k  + 1 + |m|)  \xi^k
\label{k to k+1}
\end{equation}
For Eq.\ (\ref{polynomial into eq for f}) to hold, the coefficients of $\xi^k$ must vanish for all $k$. Inserting Eq.\ (\ref{k to k+1}) into Eq.\ (\ref{polynomial into eq for f}), we thus find the following recursion relation for the coefficient $c_k$:
\begin{equation}
\frac{c_{k \! + \! 1}}{c_k} = \frac{{\displaystyle k - \frac{n}{2} \left(1 \! - \!  \frac{|m| \! + \! 1}{n} \right) + \frac{\alpha_2na}{2 \hbar^2} } }{na (k+1)(k  + 1 + |m|)}.
\label{condition on c_k's}
\end{equation}
As the polynomial in Eq.\ (\ref{polynomial}) needs to break off for finite $k$ to get a legitimate wave function (otherwise $\psi_{\xi}(\xi) \simeq e^{+\xi/2na}$ as $\xi\rightarrow\infty$, leading to a non-normalizable wave function), there must be some value $n_\xi$ for $k$ such that $c_{k \! + \! 1}=0$. Eq.\ (\ref{condition on c_k's}) tells us that this quantum number  is given by:
\begin{equation}
n_\xi = \frac{n}{2} \left(1 \! - \!  \frac{|m| \! + \! 1}{n} \right) - \frac{\alpha_2na}{2 \hbar^2}.
\label{n xi}
\end{equation}
To find the solution of  equation (\ref{psi eta eq 2}) for $\psi_\eta(\eta)$ for ${\mathcal E}=0$, we proceed in the exact same way as we did in Eqs.\ (\ref{psi xi eq 3})--(\ref{n xi}) for $\psi_\xi(\xi)$. Given its behavior at small and large $\eta$, we write $\psi_\eta(\eta)$ as (cf.\ Eq.\ (\ref{psi xi factorized}))
\begin{equation}
\psi_\eta(\eta) = \eta^{|m|/2} e^{- \eta/2na} g(\eta).
\label{psi eta factorized}
\end{equation}
We then derive an equation for $g(\eta)$, analogous to Eq.\ (\ref{eq for f}) for $f(\xi)$. The solution of this equation will be a polynomial $\sum_k \hat{c}_k \eta^k$ (where the `hat' is used to distinguish the coefficients from those in Eq.\ (\ref{polynomial}) for $f(\xi)$)  that will break off if and only if there is a value $n_\eta$ for $k$ such that (cf.\ Eq.\ (\ref{n xi})):
\begin{equation}
n_\eta = \frac{n}{2} \left(1 \! - \!  \frac{|m| \! + \! 1}{n} \right) + \frac{\alpha_2na}{2 \hbar^2}.
\label{n eta}
\end{equation}
Combining Eq.\ (\ref{n xi}) and Eq.\ (\ref{n eta}), we find
\begin{equation}
\frac{n}{2} \left(1 \! - \!  \frac{|m| \! + \! 1}{n} \right) - n_\xi = n_\eta - \frac{n}{2} \left(1 \! - \!  \frac{|m| \! + \! 1}{n} \right),
\label{n xi n eta 1}
\end{equation}
or
\begin{equation}
n = n_\xi + n_\eta + |m| + 1.
\label{n in new theory}
\end{equation}
Comparing this result in wave mechanics with the corresponding result (\ref{n in old theory}) in the old quantum theory, we notice that the difference between the two is the final term $+1$ in Eq.\ (\ref{n in new theory}). This extra term obviates the need for a special condition to rule out $|m| =0$ (cf.\ Eqs.\ (\ref{n in old quantum theory})--(\ref{n in wave mechanics}) and Fig.\ 2). As \citet[p.\ 708]{Epstein 1926} noted toward the end of his paper on the Stark effect in wave mechanics:
\begin{quotation}
\noindent
It will be remembered that the restriction for the azimuthal quantum number [$|m| > 0$] was an additional one, not following from the dynamical conditions. It was introduced by Bohr for the purpose of eliminating plane orbits, moving in which the electrons would sooner or later undergo a collusion [sic] with the nucleus. In our new theory an additional restriction is not necessary.
\end{quotation}

To conclude this section, we sketch how the formula for the energy levels in the first-order Stark effect is recovered in wave mechanics. The power-series solutions discussed above for the wave functions $\psi_{\xi}(\xi)$ and $\psi_{\eta}(\eta)$ turn out to be Laguerre polynomials. The full normalized energy eigenfunction solutions $ \psi_{n_{\xi}n_{\eta}m}(\xi,\eta,\varphi)$ for the state characterized by quantum numbers $n_{\xi},n_{\eta},m$ (with principal quantum number $n=n_{\xi}+n_{\eta}+|m|+1$) in the unperturbed case (i.e., a hydrogen atom in zero electric field) take the form
    \begin{equation}
    \label{zerofieldsols}
  \psi_{n_{\xi}n_{\eta}m} = C_{n_{\xi}n_{\eta}m}\left(\frac{\xi}{na}\right)^{|m|/2} \!\! e^{-\xi/2na}\left(\frac{\eta}{na}\right)^{|m|/2} \!\!  e^{-\eta/2na}L^{|m|}_{n_{\xi}}\left(\frac{\xi}{na}\right)
    L^{|m|}_{n_{\eta}}\left(\frac{\eta}{na}\right)e^{im\varphi},
    \end{equation}
    where $C_{n_{\xi}n_{\eta}m}$ is an explicitly known normalization constant which we will not give here (see, e.g., Condon and Shortley, 1963, p.\ 399). Introducing the Stark perturbation operator $H^{\rm Stark}\equiv \frac{1}{2}e{\mathcal E}(\xi-\eta)$,
    the standard procedure of first-order degenerate perturbation theory instructs us to calculate the matrix of the perturbing operator in the basis of the $n^2$ degenerate states of identical
    unperturbed energy corresponding to a given principal quantum number $n$. The calculation, using standard properties of Laguerre functions, gives
    \begin{equation}
    \label{1storderstarkwm}
    \langle n_{\xi}^{\prime}n_{\eta}^{\prime}m^{\prime}| \left( e{\mathcal E}\frac{\xi-\eta}{2} \right) |n_{\xi}n_{\eta}m\rangle = \frac{3}{2}e{\mathcal E}n(n_{\xi}-n_{\eta})a
    \delta_{n_{\xi}^{\prime}n_{\xi}}\delta_{n_{\eta}^{\prime}n_{\eta}}\delta_{m^{\prime}m}.
    \end{equation}
    The perturbation matrix is, in fact, diagonal in this basis, so the Stark energy shifts can be read off directly and are seen to be identical to the results obtained in the old quantum theory (see Eq.\ (\ref{Stark formula})). 

\section{Orbits versus eigenfunctions}

In this section, we show how the problematic non-uniqueness of orbits that we ran into in the old quantum theory (see section 3)  turns into the totally unproblematic non-uniqueness of bases in Hilbert space in the new quantum theory---more specifically: the non-uniqueness of bases of eigenfunctions in wave mechanics.

In wave mechanics, the stationary states are associated with eigenfunctions of the Hamilton operator of the system. Thus, to solve the problem of the hydrogen atom in spherical coordinates $(r, \vartheta, \varphi)$, we need to find normalizable (square-integrable) solutions of the time-independent Schr\"odinger equation,
\begin{equation}
-\frac{\hbar^{2}}{2\mu}\left(\frac{1}{r^{2}}\frac{\partial}{\partial r}\left(r^{2}\frac{\partial\psi}{\partial r}\right)+\frac{1}{r^{2}\sin{\vartheta}}\frac{\partial}{\partial\vartheta}\left(\sin{\vartheta}\frac{\partial\psi}{\partial\vartheta}\right)
 +\frac{1}{r^{2}\sin^{2}{\vartheta}}\frac{\partial^{2}\psi}{\partial\varphi^{2}}\right)-\frac{e^{2}}{r}\psi = E\psi,
 \label{Schpolar}
 \end{equation}
for energy eigenvalues $E<0$,  where the wavefunction $\psi$ is a function of $(r,\vartheta,\varphi)$. We highlight this choice of coordinates by using the notation $\psi^{\rm spherical}(r,\vartheta,\varphi)$ for solutions of Eq.\ (\ref{Schpolar}). The negative-energy normalizable solutions of Eq.\ (\ref{Schpolar}) correspond to the discrete energies $E_n$
labeled by the value of the principal quantum number $n$. The Schr\"odinger equation can be {\it separated} in spherical coordinates, which means that Eq.\ (\ref{Schpolar}) has solutions of the form $\psi_{n_r}(r) \psi_{n_\vartheta}(\vartheta) \psi_{n_\varphi} (\varphi)$. For each value of the principal quantum number $n = n_r + l$, the angular momentum quantum number $l = n_\vartheta$ can take on the values $0,1,2, \ldots, n-1$ and, for each value of $l$, the azimuthal quantum number $m$ (where $|m| = n_\varphi$) can take on the values $-l,-l+1, \ldots,l-1,l$. For each value of $n$, there are $\sum_{l=0}^{n-1} (2l +1) = n^2$ degenerate orthogonal solutions of Eq.\ (\ref{Schpolar}). We can conveniently label these  solutions with the values of $n$, $l$, and $m$ and introduce the notation $\psi^{\rm spherical}_{nlm}(r,\vartheta,\varphi)$ for them. Any solution of Eq.\ (\ref{Schpolar}) for $E = E_n$ must  be a linear combination of the solutions $\psi^{\rm spherical}_{nlm}(r,\vartheta,\varphi)$ with different values of $l$ and $m$ but a fixed value of $n$.

In parabolic coordinates $(\xi, \eta, \varphi)$ (see Eq.\ (\ref{parabolic coordinates})), the time-independent Schr\"odinger equation for the same system is (cf.\  Eqs.\ (\ref{Hamilton operator})--(\ref{Schroedinger eq H hydrogen})):
\begin{equation}
- \frac{\hbar^2}{2\mu} \left( \frac{4}{\xi + \eta}  \frac{\partial}{\partial \xi} \left( \xi \frac{\partial \psi}{\partial \xi} \right) +  \frac{4}{\xi + \eta}  \frac{\partial}{\partial \eta} \left( \eta \frac{\partial \psi}{\partial \eta} \right) + \frac{1}{\xi \eta} \frac{\partial^2 \psi}{\partial \varphi^2}  \right)
- \frac{2e^2}{\xi + \eta} \psi = E \psi,
\label{Schpara}
\end{equation}
where the wave function $\psi$ is now a function of $(\xi, \eta, \varphi)$. We highlight this choice of coordinates by adopting the notation $\psi^{\rm parabolic}(\xi, \eta, \varphi)$ for solutions of Eq.\ (\ref{Schpara}).\footnote{Note that only two of the three coordinates have actually been changed. The azimuthal angle coordinate $\varphi$ is the same in both coordinate systems. The $\varphi$-derivative terms in Eqs.\ (\ref{Schpolar}) and (\ref{Schpara}) are, in fact, identical, as can readily be established with the help of Eq.\ (\ref{parabolic coordinates}) for the transformation from Cartesian to parabolic coordinates.}
The Schr\"odinger equation can once again be {\it separated} in parabolic coordinates, which means that Eq.\ (\ref{Schpara}) has solutions of the form $\psi_{n_\xi} (\xi) \psi_{n_\eta} (\eta) \psi_{n_\varphi} (\varphi)$ (see Eq.\ (\ref{factorization})). For $E <0$, there are normalizable solutions only for discrete energies $E_{n}$ labeled by the principal quantum number $n = n_\xi + n_\eta + n_\varphi +1$ (see Eq.\ (\ref{n in new theory})). For any fixed value of $n$, there are $n^{2}$ degenerate orthogonal solutions, labeled by the values of the integer quantum numbers $n_\xi$, $n_\eta$, and $n_\varphi = |m|$ (where $m$ is the same as in spherical coordinates). We introduce the notation $\psi^{\rm parabolic}_{n_\xi n_\eta m}(\xi, \eta, \varphi)$ for these solutions. For a given value of $n$, all combinations of positive or zero values of $n_{\xi}$, $n_{\eta}$, and $|m|$ consistent with $n=n_{\xi}+n_{\eta}+|m|+1$ are possible. The number of such combinations is $n^2$. Any solution of Eq.\ (\ref{Schpara}) for $E = E_n$ must  be a linear combination of the solutions $\psi^{\rm parabolic}_{n_\xi n_\eta m}(\xi, \eta, \varphi)$ with different values of $n_\xi$, $n_\eta$, and $|m|$ but a fixed value of $n$.

Any solution $\psi^{\rm spherical}(r,\vartheta,\varphi)$ of Eq.\ (\ref{Schpolar})  can be immediately converted into a solution $\psi^{\rm parabolic}(\xi,\eta,\varphi)$ of Eq.\ (\ref{Schpara}) simply by expressing $(r,\vartheta,\varphi)$ in terms of $(\xi,\eta,\varphi)$. The allowed physical states are therefore identical, whichever coordinate system we use, unlike the orbits selected by the quantum conditions in the old quantum theory. However, the individual {\em separated} solutions (labeled by definite triples of quantum numbers in either coordinate system) are not in one-to-one correspondence. Since we are dealing with different representations of one and the same self-adjoint operator, any solution $\psi^{\rm spherical}_{nlm}(r,\vartheta,\varphi)$ of Eq.\ (\ref{Schpolar}) must be a linear combination of the solutions $\psi^{\rm parabolic}_{n_\xi n_\eta m}(\xi,\eta,\varphi)$ of Eq.\ (\ref{Schpara}). After all, any solution $\psi^{\rm spherical}_{nlm}(r,\vartheta,\varphi)$---with $(r,\vartheta,\varphi)$ expressed in terms of $(\xi,\eta,\varphi)$---of Eq.\ (\ref{Schpolar}) must also be a solution of Eq.\ (\ref{Schpara}) for the same energy $E=E_{n}$. The lack of a one-to-one correspondence between the separated solutions in two different coordinate systems can be regarded as the ``residue" in wave mechanics of the problem of the non-uniqueness of the orbits in the old quantum theory. This residue, of course, is no problem at all in wave mechanics.
     
The explicit separated solutions are as follows. In spherical coordinates, they are
     \begin{equation}
     \label{polarsol}
     \psi^{\rm spherical}_{nlm}(r,\vartheta,\varphi) = C_{nlm}\left(\frac{2r}{na}\right)^{l}e^{-r/na}L^{2l+1}_{n-l-1}\left(\frac{2r}{na},\right)P^{m}_{l}(\cos{\vartheta})e^{im\varphi},
    \end{equation}
while in parabolic coordinates, they are
    \begin{equation}
    \label{parasol}
    \psi^{\rm parabolic}_{n_{\xi}n_{\eta}m}(\xi,\eta,\varphi) = C_{n_{\xi}n_{\eta}m}\left(\frac{\eta\xi}{n^{2}a^{2}}\right)^{|m|/2}e^{-(\xi+\eta)/2na}L^{|m|}_{n_{\xi}}\left(\frac{\xi}{na}\right)L^{|m|}_{n_{\eta}} \left(\frac{\eta}{na}\right)e^{im\varphi},
    \end{equation}
    where $P_{l}^{m}$ and $L^{\cdots}_{\cdots}$ are the associated Legendre and Laguerre polynomials. $C_{nlm}$ and $C_{n_{\xi}n_{\eta}m}$ are normalization constants. Dropping these,
    and the identical exponential radial dependence $e^{-r/na}=e^{-(\xi+\eta)/2na}$ and azimuthal dependence $e^{im\varphi}$ in both cases, we find that 
    \begin{equation}
    \label{sepsolpolar}
    \psi^{\rm spherical}_{nlm}(r,\vartheta,\varphi) \; \propto \; r^{l}L^{2l+1}_{n-l-1}\left(\frac{2r}{na}\right)P^{m}_{l}(\cos{\vartheta}),
    \end{equation}
 and that
    \begin{equation}
    \label{sepsolpara}
    \psi^{\rm parabolic}_{n_{\xi}n_{\eta}m}(\xi,\eta,\varphi) \; \propto \; r^{|m|}\sin^{|m|}{\vartheta}L^{|m|}_{n_{\xi}}\left(\frac{\xi}{na}\right)L^{|m|}_{n_{\eta}}\left(\frac{\eta}{na}\right).
    \end{equation}
As discussed above, the functions in Eq.\ (\ref{sepsolpolar}) must be linear combinations of those in Eq.\ (\ref{sepsolpara}) (and conversely). This is tedious to demonstrate algebraically in complete generality, but easy to see for the special case of maximal azimuthal quantum number, $|m|=l$. In this case, using the addition formula
    \begin{equation}
    \sum_{n_{\xi}=0}^{n-|m|-1}L^{|m|}_{n_{\xi}}\left(\frac{\xi}{na}\right)L^{|m|}_{n-|m|-1-n_{\xi}}\left(\frac{\eta}{na}\right) = L_{n-|m|-1}^{2|m|+1}\left(\frac{\xi+\eta}{na}\right) = L_{n-l-1}^{2l+1}\left(\frac{2r}{na}\right)
    \end{equation}
    and
    \begin{equation}
     P_{l}^{l}(\cos{\vartheta}) = (2l-1)!!\sin^{l}{\vartheta},
     \end{equation}
     we find very simply that
     \begin{equation}
     \label{mequall}
     \varphi^{\rm spherical}_{nll} = (2l-1)!!\sum_{n_{\xi}=0}^{n-l-1}\varphi^{\rm para}_{n_{\xi},n-l-1-n_{\xi},m=l}.
     \end{equation}
     
For example, 
for the $2p$ states with maximal $|m|=1$, the sum in Eq.\ (\ref{mequall}) degenerates to a single term and we have a one-one correspondence between normalized  states in spherical coordinates, which we will denote as $|n \, l \, m\rangle$, and normalized  states in parabolic coordinates, which we will denote as $|n_{\xi} \, n_{\eta} \,m)$:
\begin{equation}
|2\,1 +\!1\rangle = |0\,0  +\!1), \quad |2\,1 -\!1\rangle = |0\,0  - \!1).
\end{equation}
For the $2s$ state, $m=l=0$ and the sum in Eq.\ (\ref{mequall}) contains two terms and we have
\begin{equation}
|2\;0\;0\rangle = \frac{1}{\sqrt{2}}\{|1\;0\;0) + |0\;1\;0)\}.
\label{2scomb}
\end{equation}
The remaining state ($2p$ with $m=0$) is evidently
\begin{equation}
|2\;1\;0\rangle = \frac{1}{\sqrt{2}}\{|1\;0\;0) - |0\;1\;0)\}.
\label{2pcomb}
\end{equation}
     
Once the term $eE r \! \cos{\vartheta} =\frac{1}{2}eE(\xi-\eta)$, describing an external field in the $z$-direction, is added to the Hamilton operator of the system, the problem is no longer separable in spherical coordinates, neither in the old quantum theory [$S(r,\vartheta,\varphi) \neq S_{n_r}(r) + S_{n_\vartheta}(\vartheta) + S_{n_\varphi}(\varphi)$] nor in wave mechanics [$\psi(r,\vartheta,\varphi) \neq \psi_{n_r}(r) \psi_{n_\vartheta} (\vartheta) \psi_{n_\varphi}(\varphi)$].  However, the problem continues to be separable in parabolic coordinates, both in the old quantum theory [$S(\xi, \eta, \varphi) = S_{n_\xi}(\xi) + S_{n_\eta}(\eta) + S_{n_\varphi}(\varphi)$; cf.\ Eq.\ (\ref{separation})] and in wave mechanics [$\psi(\xi, \eta, \varphi) = \psi_{n_\xi}(\xi) \psi_{n_\eta}(\eta) \psi_{n_\varphi}(\varphi)$; cf.\ Eq.\ (\ref{factorization})]. From the point of view of the old quantum theory, this means that the dynamics {\it must} be analyzed in parabolic coordinates. In wave mechanics, the separated eigenfunctions in spherical coordinates are no longer eigenfunctions of the new Hamilton operator, but  linear combinations of them give, at least to first order,  the separated eigenfunctions in parabolic coordinates! To account for the first-order Stark effect, the quantum conditions of the old quantum theory had to be imposed in parabolic coordinates. From the point of view of the new quantum theory, this is directly related to the fact that, in standard first-order degenerate perturbation theory, the matrix of the perturbing part of the Hamilton operator (here the term with the external electric field) is diagonal in the basis of the (unperturbed) states $|n_{\xi} \, n_{\eta} \,m)$ in parabolic coordinates (cf.\ Eq.\ (\ref{1storderstarkwm})). 

\section{The WKB approximation: recovering the old quantum conditions with half-integer instead of integer quantum numbers from wave mechanics}

Our analysis of the treatment of the Stark effect in hydrogen in the old quantum theory, typically hailed as one of its great successes, has turned up some serious problems. In section 3, we showed that which sets of electron orbits are allowed, even in the absence of an external electric field, depends on the coordinates in which the quantum conditions are imposed. Even though the energy levels and the level of degeneracy (both determined by the principal quantum number $n$) are the same in different coordinate systems, the eccentricities and angular momenta are not. In section 5, we saw that in wave mechanics this problem is resolved by replacing the old quantum theory's representation of physical states in terms of orbits by a representation in terms of wave functions. In view of this, it may seem almost miraculous that the old quantum theory worked even to the extent that it did.\footnote{A well-known related puzzle, which is beyond the scope of our paper, is how Sommerfeld could get the right formula for the fine-structure constant in the old quantum theory. See \citet[p. 59]{Eckert 2013} for a brief discussion of this issue. \citet{Yourgrau and Mandelstam 1979} argue that the neglect of wave-mechanical effects and the neglect of spin canceled each other out. \citet{Biedenharn 1983} takes issue with this assessment.}

This becomes even more puzzling when we consider that, in the Bohr model of the atom, the ground state of the electron (with principal quantum number $n=1$) has angular momentum $l\hbar$ with $l=n=1$, whereas in the correct theory the angular momentum of the electron is actually zero. Despite this glaring distortion of the physical situation, the theory gave the correct value for the binding energy of the electron. As we saw in sections 2 and 4, the mismatch between the angular momentum values and energy levels in the original Bohr model persisted in the old quantum theory of Bohr and Sommerfeld (see Eqs.\ (\ref{n in old quantum theory})--(\ref{n in wave mechanics}) and (\ref{n in new theory}) and Fig.\ 2), resulting in a good deal of convoluted argumentation where certain states of zero angular momentum had to be ruled ineligible on the grounds of instability. 

The seemingly accidental success of the old quantum theory can be explained on the basis of the new quantum theory. This explanation is based on an approximation scheme known as the WKB or JWKB approximation, named after \citet{Wentzel 1926b}, \citet{Kramers 1926} and \citet{Brillouin 1926}, who developed it shortly after the transition from the old to the new quantum theory in 1925--26, and \citet{Jeffreys 1924} who developed it right before that watershed in a different context.\footnote{See \citet[pp.\ 8--9 and pp.\ 20--36]{Mehra Rechenberg} for a brief history of the (J)WKB method and references to the most important contributions to the further development of the method. For additional references, see the bibliographies of two monographs on the subject, \citet{Heading 1962} and \citet{Froeman and Froeman 1965}. For a  textbook treatment of WKB, see, e.g., \citet[pp.\ 131--138, sec.\ 25, ``The phase integral approximation"]{Stehle 1966}. \label{WKB refs}} 

The WKB method, as we will refer to it hereafter, can be seen as providing a ``halfway house" between the old and the new theory. It amends the quantum conditions of the old quantum theory, replacing many integer quantum numbers of the original theory by half-integer quantum numbers. In fact, before the advent of the new quantum mechanics, several physicists, including Reiche and Pauling, had made use of half-integer quantum numbers, which typically gave better results than integer ones (e.g., in the case of the specific heat of hydrogen), though in at least one case (that of the electric susceptibility of some diatomic gases) it only made matters worse.\footnote{See \citet[especially pp.\ 158, 166]{Gearhart 2010} for the case of Reiche and specific heat and \citet{Midwinter and Janssen 2013} for the case of Pauling and electric susceptibilities (see p.\ 186 for an explanation of why in this case half-integer quantum numbers only made matters worse). In both cases, the quantization rule for angular momentum in the old quantum theory, $L = l \hbar$ was changed to $L = (l + \frac{1}{2}) \hbar$. \citet{Kramers 1926} explicitly mentioned  half-integer quantization in his paper on the new approximation scheme. However,  as we will see, the half-quantum numbers in the two examples mentioned above are not the half-quantum numbers dropping out of the WKB scheme (see Eq.\ (\ref{Langer}) below). Other instances of factors of $ \frac{1}{2}$ cropping up in the late teens and early twenties turned out to be related to spin and have no connection to WKB either. However, the $ \frac{1}{2}h \nu$ term in the energy $E = h\nu (n + \frac{1}{2})$ of the harmonic oscillator, i.e., its zero-point energy, is reproduced by the WKB approximation scheme. \label{half-integer examples}}  With the WKB amendment of the quantum conditions of the old quantum theory, the connection between the values of the energy and angular momentum is the same as in the new quantum theory (e.g., Eq.\ (\ref{n in old quantum theory}), $n = n_\xi + n_\eta + |m|$, changes to Eq.\ (\ref{n in wave mechanics}), $n = n_\xi + n_\eta + |m| +1$). As we will see in this section, the WKB approximation is key to understanding the remarkable, if partial, success of the old quantum theory.

The first inklings of the semi-classical analysis that would eventually evolve into the WKB method can be found in a heuristic
argument that \citet{Schroedinger 1926a} used in his first paper on wave mechanics. This is especially clear in the case of \citet{Wentzel 1926b} and \citet{Brillouin 1926}, who both took the classical Hamilton-Jacobi equation as the starting point of their analysis. \citet{Kramers 1926} only made the connection between his approximation scheme for the Schr\"odinger equation and the classical Hamilton-Jacobi equation explicit in sec.\ 3 of his paper, in which he compared his approach to that of Brillouin and Wentzel (ibid., pp.\ 834--836). 

Schr\"odinger showed that the reformulation of the classical Hamilton-Jacobi equation in terms of a variational principle leads to the Schr\"odinger equation if we demand that the functional being varied be extremal rather than simply zero, as in the classical case. The connection requires that we express the wave-function $\psi(x)$ in terms of a phase function $S(x)$, which is complex though its leading term (which does not contain $\hbar$) is real:
 \begin{equation}
 \psi(x) \equiv  \exp{\! \left(\frac{i}{\hbar}S(x)\right)}.
 \label{wkb1}
 \end{equation}
As the notation suggests and as will become clear below, the phase function $S(x)$ is closely related to Hamilton's principal function $S(x)$ in Hamilton-Jacobi theory.

We will only consider systems in one dimension. This restriction is less constraining than it may seem, as the success of the Bohr-Sommerfeld approach depended on the existence of coordinate systems in which the three-dimensional classical motion could be separated into three effectively independent one-dimensional motions. 

Consider a Hamiltonian of the form $H= p^2/2\mu + V(x)$. Replacing $p$ by $(\hbar/i) \, d/dx$, we find the Schr\"odinger equation in one dimension\footnote{As we saw in section 4 (see especially Eq.\ (\ref{p to operator p})), this substitution is similar to the replacement of $p$ by $dS/dx$ in Hamilton-Jacobi theory.}
\begin{equation}
\left( - \frac{\hbar^2}{2 \mu} \frac{d^2}{dx^2} + V(x) \right) \psi(x) = E \psi(x).
\end{equation}
This can be rewritten as
    \begin{equation}
    \label{1DSchro}
    \psi^{\prime\prime}(x)+\frac{2\mu}{\hbar^{2}}(E-V(x)) \, \psi(x)=0.
    \end{equation}
We now insert the expression for $\psi(x)$ in Eq.\ (\ref{wkb1}), using that
\begin{equation}
\psi^\prime(x) = \frac{i}{\hbar} S^\prime(x) \psi(x),
\quad \quad
\psi^{\prime \prime}(x) = \left( \frac{i}{\hbar} S^{\prime \prime}(x) - \frac{1}{\hbar^2} S^{\prime}(x)^2 \right) \psi(x),
\end{equation}
and divide by $\psi(x)$:
\begin{equation}
\frac{i}{\hbar} S^{\prime \prime}(x) - \frac{1}{\hbar^2} S^{\prime}(x)^2 + \frac{2 \mu}{\hbar^2} (E - V(x)) = 0.
\end{equation}
Multiplying by $-\hbar^2$ and rearranging terms, we find
\begin{equation}
S^{\prime}(x)^2 - 2 \mu( E - V(x)) - i\hbar S^{\prime \prime}(x) =0.
\label{class HJ eq + hbar term}
\end{equation}
We recover this equation---{\it except} for the term depending on $\hbar$---if we substitute $S^\prime(x)$ for $p$ in the equation $H(p, x) = E$, with  $H= p^2/2\mu + V(x)$. In other words, if the term with $\hbar$ is neglected, Eq.\ (\ref{class HJ eq + hbar term}) is just the classical Hamilton-Jacobi equation for the system.\footnote{This relation between the Schr\"odinger equation and the classical Hamilton-Jacobi equation is emphasized, for instance, in the opening paragraph of the paper in which \citet{Jordan 1927} introduced transformation theory.} 

We introduce the quantities
\begin{equation}
f(x) \equiv S^\prime(x)
\label{def f}
\end{equation}
and 
\begin{equation}
p(x) \equiv \sqrt{2 \mu (E - V(x))}.
\label{def p}
\end{equation}
Note that $2 \mu (E - V(x))$ gives the square of the momentum as a function of $x$. Also note that $p$ is real only if $E > V(x)$. Substituting Eqs.\ (\ref{def f}) and (\ref{def p}) into Eq.\ (\ref{class HJ eq + hbar term}), we find
\begin{equation}
f(x)^2 - p(x)^2 - i \hbar f^\prime(x) = 0.
\label{wkb3}
\end{equation}
Treating $\hbar$ as a small parameter, we can solve Eq.\ (\ref{wkb3}) iteratively. Using that $f(x) = \pm p(x)$ to zeroth order, we obtain at the next level of approximation:
\begin{equation}
f^{2} = p(x)^{2} \pm i\hbar p^\prime(x).
\label{wkb3a}
\end{equation}
Squaring and then differentiating both sides of Eq.\ (\ref{def p}), we see that $2p(x) p^\prime(x) = - 2 \mu V^\prime(x)$ or that
\begin{equation}
p^\prime(x) = - \frac{\mu V^\prime(x)}{p(x) }.
\label{wkb3b}
\end{equation}
Using this relation, we can rewrite Eq.\ (\ref{wkb3a}) as
\begin{equation}
f^{2} = p(x)^{2} \mp  \frac{i\hbar \mu V^\prime(x)}{p(x)}.
\label{wkb4}
\end{equation}
It follows that, to first order in $\hbar$,
\begin{eqnarray}
f(x) &=& \sqrt{p(x)^{2} \mp  \frac{i\hbar \mu V^\prime(x)}{p(x)}} \nonumber \\
&=& \pm p(x) \left( 1\mp \frac{i\hbar \mu V^\prime(x)}{2p(x)^3} \right) \label{wkb7} \\
&=& \pm p(x) -\frac{i\hbar \mu V^\prime(x)}{2p(x)^2}. \nonumber
\end{eqnarray}
The semi-classical approximation that we are considering here is valid only if the second term on the right-hand-side of (\ref{wkb7}) is small compared to the first, i.e., if 
 \begin{equation}
\left| \frac{ \hbar \mu V^{\prime}(x)}{ 2 p(x)^{3}} \right| < \!\! < 1.
 \label{wkb8}
 \end{equation}
 This condition may look peculiar at first sight but it  amounts to the perfectly natural requirement that the relative change in the ``local" momentum $p(x)$ (due to the variation of the potential) over a single de Broglie wavelength be small \citep[p.\ 131]{Stehle 1966}. The ``local" de Broglie wavelength $\lambda(x)$ is given by $h/p(x)$. The relative change in $p(x)$ over this distance can be written as
\begin{equation}
\frac{\delta p(x)}{p(x)} = \frac{p^{\prime}(x)\lambda(x)}{p(x)} = - \frac{ \hbar \mu V^{\prime}(x)}{p(x)^{3}},
\label{wkb8a}
\end{equation}
where we used Eq.\ (\ref{wkb3b}) for $p^\prime(x)$. Comparing Eqs. (\ref{wkb8}) and (\ref{wkb8a}), we see that the condition for the validity of our semi-classical approximation is just that $|\delta p(x)| < \!\! < |p(x)|$. 
     
In order to find the Schr\"odinger wave function $\psi(x)$ in Eq.\ (\ref{wkb1}) in this semi-classical approximation, we need to integrate Eq.\ (\ref{wkb7}) to get from the derivative $S^\prime(x) \equiv f(x)$ of the phase factor in the wave function to this phase factor $S(x)$ itself:
\begin{equation}
 S(x) = \pm \! \int^{x} \!\! p(y)dy - \frac{i\hbar \mu}{2} \! \int^x \! \frac{V^\prime(y)}{p(y)^2} dy.
\label{swkb1}
\end{equation}
Using Eq.\ (\ref{def p}) for $p(y)$ in the second integral, we can rewrite this as
\begin{equation}
S(x) = \pm \! \int^{x} \!\! p(y)dy - \frac{i\hbar }{4} \! \int^x \! \frac{V^\prime(y)}{E - V(y)} dy.
\label{swkb1a}
\end{equation}
The second integral is equal to
\begin{equation}
- \ln{(E - V(x))} = - \ln{ \left( \frac{p(x)^2}{2 \mu} \right) } = - 2 \ln{p(x)} + \ln{2 \mu}.
\end{equation}
Substituting this expression into Eq.\ (\ref{swkb1a}), we find
\begin{equation}
S(x) = \pm \! \int^{x} \!\! p(y)dy + \frac{i\hbar }{2} \ln{p(x)} + {\rm constant}.
\label{swkb2}
\end{equation}
Inserting this result into the expression for the Schr\"odinger wave function $\psi(x)$ in Eq.\ (\ref{wkb1}), one finds, apart from (at this point uninteresting) normalization factors:
     \begin{equation}
     \label{wfwkb1}
     \psi(x) \propto \frac{1}{\sqrt{p(x)}} \exp{\! \left(\pm \frac{i}{\hbar}\int^{x} \!\! p(y)dy \right)}.
     \end{equation}
In general, the wave-function will be a linear combination of the two possibilities for $\psi(x)$ in Eq.\ (\ref{wfwkb1}):
     \begin{equation}
     \psi(x) = \frac{1}{\sqrt{p(x)}} \left( A \, \exp{\! \left( +\frac{i}{\hbar}\int^{x} \!\!  p(y)dy \right)}+B \, \exp{\! \left(-\frac{i}{\hbar}\int^{x} \!\!  p(y)dy\right)}\right),
     \label{wfwkb2}
     \end{equation}
     with the (in general complex) coefficients $A$ and $B$ and the lower bounds of the integrals (not shown here) chosen to respect the relevant physical boundary conditions.
     
     The preceding analysis should be regarded as holding in the region in which $E>V(x)$ so that $p(x) = \sqrt{2\mu (E - V(x))}$ (see Eq.\ (\ref{def p})) is real. In the region where $E<V(x)$, which is classically forbidden but critical in quantum mechanics, it is more convenient to write Eq.\ (\ref{wfwkb2}) in terms of $\pi(x) \equiv \sqrt{2 \mu (V(x)-E)}$, which is real when $E<V(x)$. Since $p(x) = i \pi(x)$, the exponentials in Eq.\ (\ref{wfwkb2}) now are real:
     \begin{equation}
   \psi(x) = \frac{1}{\sqrt{\pi(x)}} \left(C \, \exp{\! \left(- \frac{1}{\hbar}\int^{x} \!\! \pi(y)dy \right)}+D \, \exp{\! \left(+ \frac{1}{\hbar}\int^{x} \!\! \pi(y)dy \right)} \right).
     \label{wfwkb3}
     \end{equation}
Once again, the (complex) coefficients $C$ and $D$ (and the lower bounds of the integrals) must be chosen so that the wave function has the right asymptotic behavior (e.g., vanishes appropriately at spatial infinity).
     
\begin{figure}[h]
   \centering
   \includegraphics[width=4in]{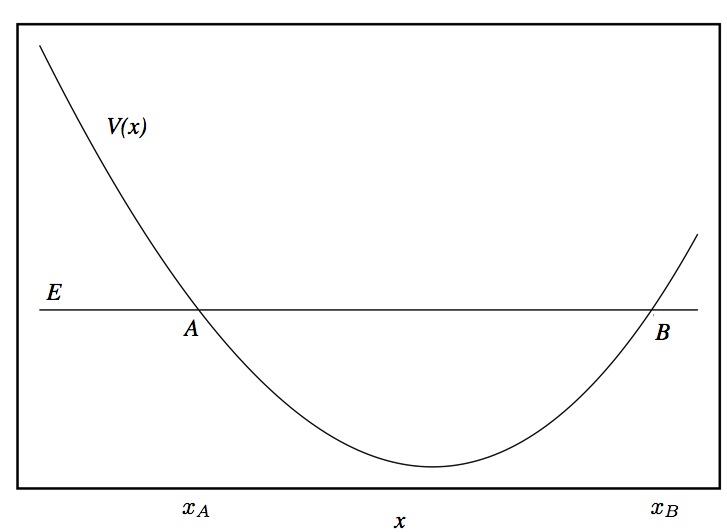} 
   \caption{{\small Potential energy plot for a classically bound particle.}}
   \label{wkbpot}
\end{figure}

The semi-classical approximations to the wave function  in Eqs.\ (\ref{wfwkb2}) and (\ref{wfwkb3})  cannot be expected to hold at points where $p(x) \approx 0$, i.e., near the classical turning points where $E=V(x)$. This can be seen upon inspection of the original 
expansion in Eq.\ (\ref{wkb7}).  Consider, for example, the case of a classically bound particle of energy $E$ moving in the potential shown in Fig.\ \ref{wkbpot}. The particle executes oscillatory motion between the points $x_{A}$ and $x_{B}$, coming to a stop at precisely those points. In wave mechanics, the de Broglie wavelength goes to infinity when the momentum vanishes, so the assumption that the potential only varies slightly over a single wavelength must certainly fail at such points. Instead, a more careful analysis of the original Schr\"odinger equation in the  immediate neighborhood of the turning point(s)
shows that the solutions take the form of Airy functions (Bessel functions of order 1/3). These solutions can then be extended to a form that matches the  semi-classical forms given in Eqs.\ (\ref{wfwkb2}) and (\ref{wfwkb3}) that are correct far from the turning point(s). The details of this procedure are irrelevant for our purposes. We should emphasize, however, that this is the central technical problem facing the WKB approach. Without a careful treatment of the solution in the neighborhood of turning points, essential features of the semi-classical quantization are missed.\footnote{For a careful treatment, see, e.g., \citet[Vol.\ 2, pp.\ 1092--1099]{Morse and Feshbach} and the monographs mentioned in note \ref{WKB refs}.} 

We examine the behavior of the WKB solution for a bound state as we go from $x = - \infty$ to $x = + \infty$ (cf.\ Fig.\ \ref{wkbpot}). For $x  <\!\!< x_A$, the solution has the form of Eq.\ (\ref{wfwkb3}), with the lower bound of the integrals chosen as $x_{A}$ and the constant $C$ set to zero. The term involving the coefficient $D$ will vanish exponentially as $x\rightarrow -\infty$ (as the exponent is the integral of a positive quantity from $x_A$ to $x$). The extension of the appropriate combination of Airy functions to the region to the right of the turning point $x_A$ then yields an oscillatory solution of the form of Eq.\ (\ref{wfwkb2}). We must here assume that we are able to go far to the right (many wavelengths) of $x_{A}$, while remaining to the left of the turning point $x_{B}$. The specific form of this solution  (the overall constant is irrelevant) is (see, e.g., Kramers, 1926, p.\ 831, Eq.\ 12)
 \begin{equation}
       \label{xasol}
       \psi(x) \propto  \cos{\left( \frac{1}{\hbar} \int_{x_{A}}^x \!\! p(y)dy -\frac{\pi}{4}\right)},\quad x_{A} < \!\!< x <\!\!< x_{B},
     \end{equation}
where the much-greater-than signs imply many wavelengths.
     
The same procedure can be applied at the right turning point $x_B$. For $x >\!\!> x_B$, the solution once again has the form of Eq.\ (\ref{wfwkb3}), now with the lower bound of the integrals chosen as $x_B$ and the constant $D$ set to zero. In this case, the term involving the coefficient $C$ will  vanish exponentially as $x\rightarrow \infty$. Once again, we need to connect the exponentially decreasing solution for $x>\!\!>x_{B}$
to an oscillatory solution to the left of $x_{B}$ similar to the one in Eq.\ (\ref{xasol}) (not surprisingly, from the symmetry of the situation)
          \begin{equation}
       \label{xbsol}
       \psi(x) \propto  \cos{\left( \frac{\pi}{4}- \frac{1}{\hbar} \int_{x}^{x_{B}} \!\! p(y)dy \right)},\quad x_{A} < \!\!< x <\!\!< x_{B}.
     \end{equation}
     As the solutions in Eqs.\ (\ref{xasol}) and (\ref{xbsol}) must agree, the arguments of the cosine in these two equations must be equal to one another {\it modulo} $n \pi$, where $n$ is an arbitrary integer:
     \begin{equation}
     \label{wkb10}
      \frac{1}{\hbar}  \int_{x_{A}}^{x} \!\! p(y)dy  -\frac{\pi}{4} =  \frac{\pi}{4}- \frac{1}{\hbar}  \int_{x}^{x_{B}} \!\! p(y)dy  +n\pi.
      \end{equation}
  Rearranging and renaming the integration variable, we can rewrite this as
     \begin{equation}
     \label{wkbqu1}
     \frac{1}{\hbar}\int_{x_{A}}^{x_{B}}p(x)dx = \left(n+ {\textstyle \frac{1}{2}} \right)\pi.
     \end{equation}
    Taking an integral over a complete cycle of the classical motion,
     from $x_{A}$ up to $x_{B}$ and back again, we find:
          \begin{equation}
     \label{wkbqu2}
       \oint p(x)dx = 2\int_{x_{A}}^{x_{B}}p(x)dx = \left(n+ {\textstyle \frac{1}{2}} \right) 2 \pi \hbar = \left(n+ {\textstyle \frac{1}{2}} \right)h.
     \end{equation}
     This is precisely the quantum condition  of the old quantum theory (cf.\ Eqs.\ (\ref{quantum conditions})--(\ref{I-phi}))
     {\em with the critical addition of the $\frac{1}{2}$ piece} (cf.\ note \ref{half-integer examples}).
    
 Of course, the conditions of validity of the derivation really require $n>\!\!>1$, as the left-hand-side of Eq.\ (\ref{wkbqu1}) is just the phase change of the wave function between the turning points. Nevertheless, for the special case of the Coulomb potential, this extra term---now referred to as a {\em Maslov index} \citep[p.\ 211]{Gutzwiller 1990}---is precisely what is needed to restore the consistency between the angular momentum and energy quantum numbers lost in the old quantum theory.\footnote{As mentioned above (see note \ref{half-integer examples}), the extra term in the quantum condition (\ref{wkbqu2}) can also be used to justify the zero-point energy term of the harmonic oscillator. It changes the quantum condition on the energy of the harmonic oscillator from $E = nh\nu$ to $E = (n + \frac{1}{2}) h\nu$.} 
     
The improvement in the situation is seen most directly if we impose the quantum conditions of the old quantum theory in parabolic coordinates in the case of the hydrogen atom. The motion in the coordinates $\xi$ and $\eta$ involve the Coulomb potential and each have turning points qualitatively of the form indicated in Fig.\ \ref{wkbpot}. In section 2, we labeled these turning points $(\xi_{\rm min}, \xi_{\rm max})$ and $(\eta_{\rm min}, \eta_{\rm max})$, respectively (see Eqs. (\ref{roots xi}) and (\ref{roots eta})). The WKB amendment of the quantum conditions of the old quantum theory requires that we replace $n_{\xi}$ and $n_{\xi}+\frac{1}{2}$ in Eq.\ (\ref{I-xi}) for the action variable $I_\xi$ and, likewise, that we replace $n_{\eta}$ by $n_{\eta}+\frac{1}{2}$ in Eq.\ (\ref{I-eta}) for the action variable $I_\eta$. The third coordinate, the azimuthal angle $\varphi$, does not contain the potential. The motion in this coordinate corresponds to free angular motion.
 The wave function must be single-valued in $\varphi$, so the solutions $e^{2\pi in_{\varphi}\varphi}$ require that the corresponding action variable (see Eq.\ (\ref{I-phi})) 
     \begin{equation}
     I_{\varphi} = \oint p_{\varphi}d\varphi = n_{\varphi}h,
     \end{equation}
(with $n_{\varphi}$ and integer) lack the extra term of $\frac{1}{2}$. The difference in character  between the motion in the coordinates $\xi$ and $\eta$ and the motion in the coordinate $\varphi$ has a classical correlate in the distinction between {\it libration} and {\it rotation} \citep[p.\ 453, Fig.\ 10.2]{Goldstein}.

The upshot then is that the relation between the principal quantum number $n$ and the quantum numbers $n_\xi$, $n_\eta$, and $n_\varphi = |m|$ in the old quantum theory changes from
\begin{equation}
n = n_{\xi}+n_{\eta}+ |m|
\end{equation}
(see Eq.\ (\ref{n in old theory})) to
\begin{equation}
n = n_{\xi}+n_{\eta}+ |m| +1,
\end{equation}
which is precisely the result obtained in wave mechanics (see Eq.\ (\ref{n in new theory}); see also Eqs.\ (\ref{n in old quantum theory}) and (\ref{n in wave mechanics})). The contorted reasoning needed in the old quantum theory to remove certain states with $m=0$ is  rendered moot.
     
The efficacy of the WKB treatment in bringing the quantum conditions of the old quantum theory into closer compliance with the results of wave mechanics is  not restricted to the use of parabolic coordinates. However, the analysis for the more commonly used polar coordinates requires, somewhat unexpectedly, a more careful treatment than that given above for parabolic coordinates. One result of this analysis is a clearer understanding of the semi-classical ``Langer modification" \citep{Langer 1937} in which the centrifugal term in the radial Schr\"odinger equation is changed in the following way:\footnote{This modification explains the efficacy of the introduction of half-quantum numbers in several instances in the early 1920s (cf. note \ref{half-integer examples}).}
     \begin{equation}
     \label{Langer}
     \frac{\hbar^{2}l(l+1)}{2\mu r^{2}} \rightarrow \frac{\hbar^{2}(l+\frac{1}{2})^{2}}{2\mu r^{2}}.
     \end{equation}
We  conclude  our discussion of the WKB approach with a brief discussion of its application in this case (for more details, see, e.g., Morse and Feshbach, 1953, Vol.\ 2, p.\ 1101). The starting point is the familiar radial Schr\"odinger equation in a central potential, where, for simplicity, we take the Coulomb potential $V(r)=- 2/ar$ of the hydrogen atom (where $a \equiv \hbar^{2}/ \mu e^{2}$ is the Bohr radius) :
    \begin{equation}
    \label{radSchro1}
    \frac{d^{2}R(r)}{dr^{2}} + \left(\frac{2 \mu E}{\hbar^{2}} + \frac{2}{ar} -\frac{l(l+1)}{r^{2}} \right) R(r) =0.
    \end{equation}
We introduce some substitutions to transform this equation into one of the form of the  one-dimensional Schr\"odinger equation (\ref{1DSchro}), which we analyzed at the beginning of this section. First, we define a rescaled energy
\begin{equation}
-\kappa^{2} \equiv \frac{2 \mu E}{\hbar^{2}}
\label{definition kappa}
\end{equation}
(the negative sign as we are considering bound-state solutions only). More importantly, we introduce a rescaled dimensionless coordinate variable $x$ defined by 
\begin{equation}
r=ae^{x}.
\label{definition r}
\end{equation}
The range of this new variable $x$ is $-\infty<x<+\infty$, just as in the example discussed above (cf.\ Fig.\ \ref{wkbpot}), whereas the range of the original variable $r$ is $0 < r < \infty$. As explained by \citet[p.\ 674]{Langer 1937}, the WKB approximation method fails for Eq.\ (\ref{radSchro1}) because of difficulties one runs into when examining the behavior of the solution at $r=0$. Finally, we redefine the dependent variable by introducing a new one-dimensional wave function $\psi(x)$ given by
    \begin{equation}
    \label{radSchro2}
    R(r) \equiv e^{x/2}\psi(x).
    \end{equation}
With the help of Eqs.\ (\ref{definition r}) and (\ref{radSchro2}), $R^{\prime \prime}(r)$ in Eq.\ (\ref{radSchro1}) can be expressed in terms of $\psi(x)$ and its second-order derivative $\psi^{\prime\prime}(x)$. From Eq.\ (\ref{definition r}) it follows that $x = \ln{(r/a)}$ and that 
\begin{equation}
\frac{d}{dr} = \frac{dx}{dr} \,  \frac{d}{dx} = \frac{1}{r} \frac{d}{dx} = \frac{1}{a} \, e^{-x}  \frac{d}{dx}.
\end{equation}
Hence, $R^\prime(r)$ can be written as
\begin{eqnarray}
R^\prime(r) & = &  \frac{1}{a} \, e^{-x}   \frac{d}{dx} \left\{ \, e^{x/2}\psi(x) \right\} \nonumber \\
 & & \label{R prime} \\
 & = &  \frac{1}{a} \, e^{-x/2}  \left\{ \, {\textstyle \frac{1}{2}} \psi(x) + \psi^\prime(x) \right\}, \nonumber
\end{eqnarray}
and $R^{\prime \prime}(r)$ as
\begin{eqnarray}
R^{\prime\prime}(r) & = &  \frac{1}{a^2} \, e^{-x}   \frac{d}{dx}  \left\{ \, e^{-x/2} \left( {\textstyle \frac{1}{2}} \psi(x) + \psi^\prime(x) \right) \right\} \nonumber \\
 & = &  \frac{1}{a^2} \, e^{-3x/2}  \left\{  {\textstyle - \frac{1}{2}} \left[ {\textstyle \frac{1}{2}} \psi(x) + \psi^\prime(x) \right] + {\textstyle \frac{1}{2}} \psi^\prime(x) + \psi^{\prime\prime}(x) \right\}  
 \label{R double prime} \\
 & = & \frac{1}{a^2} \, e^{-3x/2} \left\{ \psi^{\prime\prime}(x) - {\textstyle \frac{1}{4}} \psi(x) \right\}.
 \nonumber
\end{eqnarray}
Inserting Eqs.\ (\ref{definition kappa})--(\ref{radSchro2}) and (\ref{R double prime}) into Eq.\ (\ref{radSchro1}), we find:
\begin{equation}
\frac{1}{a^2} \, e^{-3x/2} \left\{ \psi^{\prime\prime}(x) - {\textstyle \frac{1}{4}} \psi(x) \right\} 
+ \left( -\kappa^2 + \frac{2}{a^2 e^x} - \frac{l(l+1)}{a^2 e^{2x}} \right) e^{x/2} \psi(x) = 0.
\label{radSchro2a}
\end{equation}
If this equation is multiplied by $a^2 e^{3x/2}$, it reduces to
\begin{equation}
\psi^{\prime\prime}(x) + \left(  -\kappa^2 a^2  e^{2x} + 2 e^x - (l+ {\textstyle \frac{1}{2}})^2  \right) \psi(x) = 0.
\label{radSchro2b}
\end{equation}
Note that the terms with $- \frac{1}{4} \psi(x)$ and $l(l+1) \psi(x)$ in Eq.\ (\ref{radSchro2a}) combine to give the term $(l+ \frac{1}{2})^2 \psi(x)$ in Eq.\ (\ref{radSchro2b}). This is  the ``Langer modification" mentioned above (see Eq.\ (\ref{Langer})).  

In going from Eq.\ (\ref{radSchro1}) for $R(r)$ (with $0 < r < +\infty$) to Eq.\ (\ref{radSchro2b}) for $\psi(x)$ (with $- \infty < x < + \infty$), we have transformed the radial  Schr\"odinger equation into an equation of the form of Eq.\ (\ref{1DSchro}). We can thus apply the same WKB techniques that we applied to Eq.\ (\ref{1DSchro}) to Eq.\ (\ref{radSchro3}). 

It will be to convenient introduce the rescaled binding energy ${\mathcal E} \equiv \kappa^{2}a^{2}$ (not to be confused with the strength of the external electric field for which we used the notation ${\mathcal E}$ above). This quantity takes on the values $1/n^2$ for the quantized energy levels of the hydrogen atom, where $n=1,2,3, \dots$ is the principal quantum number. Eq.\ (\ref{radSchro2b}) thus becomes
\begin{equation}
\psi^{\prime\prime}(x) + \left( -{\mathcal E}  e^{2x} + 2 e^x - (l+ {\textstyle \frac{1}{2}})^2  \right) \psi(x) = 0.
\label{radSchro3}
\end{equation}
Just as we defined the momentum $p(x)^2$ in Eq.\ (\ref{def p}) as $\hbar^2$ times the factor multiplying $\psi(x)$ in Eq.\ (\ref{1DSchro}), we now define $p(x)^2$ as $\hbar^2$ times the factor multiplying $\psi(x)$ in Eq.\ (\ref{radSchro3})
     \begin{equation}
     \label{radmom}
     p(x)^{2} \equiv \hbar^{2} \left( - {\mathcal E}e^{2x}+2e^{x}-(l+{\textstyle \frac{1}{2}})^{2} \right).   
     \end{equation}
As in the case of Eq.\ (\ref{1DSchro}), this problem corresponds to a classical motion between turning points $x_{\rm min}$ and $x_{\rm max}$ (cf.\ Fig.\ \ref{wkbpot}). It is easier to find $x_{\rm min}$ and $x_{\rm max}$ if we switch from $x$ to $\rho \equiv e^x$ and determine $\rho_{\rm min} \equiv e^{x_{\rm min}}$ and $\rho_{\rm max} \equiv e^{x_{\rm max}}$ instead.  In terms of $\rho$, Eq.\ (\ref{radmom}) becomes 
\begin{equation}
\frac{p(\rho)^{2}}{\hbar^{2}} \equiv  - {\mathcal E} \rho^2 +2 \rho -(l+{\textstyle \frac{1}{2}})^{2}.
\label{radmom1}
\end{equation}
The turning points occur at the roots  $\rho_{\rm min}$ and $\rho_{\rm max}$  of the quadratic equation obtained by setting the right-hand side of Eq.\ (\ref{radmom1}) equal to zero.\footnote{For these roots to be real, the discriminant condition requires $\frac{1}{\mathcal E} = n^{2} > (l+ {\textstyle \frac{1}{2}})^{2}$, which is obviously true for the hydrogen atom (see, e.g., Eq.\ (\ref{wkbrad5}) below).} The right-hand side of Eq.\ (\ref{radmom1}) can be written in the form ${\mathcal E} (\rho - \rho_{\rm min}) (\rho_{\rm max} - \rho)$, where
\begin{equation}
\rho_{\rm max}+\rho_{\rm min} = 2/\mathcal{E}, \quad \quad \rho_{\rm max}\rho_{\rm min} = (l+ {\textstyle \frac{1}{2}})^{2}/\mathcal{E}
\label{roots rho}
\end{equation}
(cf.\ Eqs. (\ref{roots xi})--(\ref{a & b})). The WKB quantization rule (\ref{wkbqu1}) therefore gives
     \begin{eqnarray}
     \label{wkbrad1}
     \frac{1}{\hbar}\int_{x_{\rm min}}^{x_{\rm max}}p(y)dy &  =& \sqrt{\mathcal{E}}\int_{\rho_{\rm min}}^{\rho_{\rm max}} \!\! \sqrt{(\rho-\rho_{\rm min})(\rho_{\rm max}-\rho)} \; \frac{d\rho}{\rho} \nonumber \\
     & & \label{wkbrad2} \\
         &=& \sqrt{\mathcal{E}}\;\frac{\pi}{2} \left(\,\rho_{\rm max}+\rho_{\rm min}-2\sqrt{\rho_{\rm max}\rho_{\rm min}} \,\right) = (n_{r}+ {\textstyle \frac{1}{2}})\pi, \nonumber
     \end{eqnarray}
 where we used Eq.\ (\ref{standard integral 1}) to evaluate the integral. With the help of Eq.\ (\ref{roots rho}), we can rewrite the last equation as
     \begin{equation}
     \label{wkbrad4}
      {\textstyle \frac{1}{2}} \left( (2/\sqrt{\mathcal{E}}) -2(l+ {\textstyle \frac{1}{2}}) \right) = (n_{r}+ {\textstyle \frac{1}{2}}).
     \end{equation}
Substituting the principal quantum number $n$ for $1/\sqrt{\mathcal{E}}$ and rearranging terms, we obtain:
\begin{equation}
 n =  (n_{r}+ {\textstyle \frac{1}{2}}+l+ {\textstyle \frac{1}{2}}) = n_{r}+l+1.
\label{wkbrad5}
\end{equation}
This agrees exactly with the expression for the principal quantum number in terms of the radial and angular quantum numbers in the new theory.\footnote{The peculiar character of the Coulomb potential in conveniently yielding the correct values for the quantized energies in the {\em first} non-trivial order of the semi-classical expansion has been examined in detail by \citet{Hainz and Grabert 2011}, who show explicitly why the semi-classical energy levels obtained above remain untouched, as we know they must, in higher orders in a systematic expansion in powers of Planck's constant.} Just as in the case of parabolic coordinates, the needed extra term of $+1$ arises from two separate terms of $\frac{1}{2}$ arising from a careful WKB analysis of the solutions. The non-zero Maslov index appearing in the radial quantization integral in Eq.\ (\ref{wkbrad4}) has an interesting consequence in terms of an orbit interpretation: the possibility of a circular orbit, with degenerate roots $\rho_{\rm max}=\rho_{\rm min}$ (and therefore vanishing radial action integral) is removed. Even in the case of the ground state, with $n_{r}=0,n=1,l=0$, the associated classical motion involves a libration between two distinct radial distances $r = (1 \pm  \sqrt{3}/2)a$. Such a motion, with zero angular momentum, is  classically impossible, of course, but we have left the realm of the classical Hamilton-Jacobi equation by including (in a limited way) quantum-mechanical effects.

\section{Conclusion: Stark contrasts between the old and the new quantum theory}

In the mid-1920s, a number of physicists working on the old quantum theory became increasingly suspicious of the notion of electron orbits. In the transition from the old quantum theory to both matrix mechanics and wave mechanics, orbits were discarded altogether (see \citet{Duncan and Janssen 2007} for the case of matrix mechanics). What we have shown in this paper, using the account of the Stark effect in the old and the new quantum theory as a striking example, is that orbits had become highly problematic well before the developments of the mid-1920s. The most important theorists working on the Stark effect---Sommerfeld, Epstein, and Kramers (Schwarzschild died shortly after making his contribution)---were well aware of these difficulties, especially of the two problems that we focused on in this paper. Yet, they offered only stopgap solutions for one of these problems and essentially ignored the other. 

The two problems were the following. First, the relation between the quantum numbers for energy and angular momentum was such that the basic quantum conditions had to be supplemented by some ad-hoc extra conditions to rule out various pathological orbits (see section 2). Second, the orbits selected by the quantum conditions of the old quantum theory depend on the coordinates in which these conditions are imposed (see section 3). 

The new quantum theory  avoids both problems (see section 4).  In sections 5 and 6, we used the new theory to elucidate the problems in the old one. The first problem could be fixed systematically with the help of the new theory (see section 6). The new theory, however, also makes it painfully clear that the second problem just goes to show that orbits are ill-suited to represent physical states in atomic physics.

The problematic relation between the quantum numbers for energy and angular momentum could be fixed by amending the basic quantum conditions of the old quantum theory, replacing many but not all integer quantum numbers by half-integer ones. In section 6, we reviewed how these amendments can be derived using the WKB approximation method for finding approximate semi-classical solutions of the Schr\"odinger equation. If the motion in a particular coordinate is a libration, half-integer quantum numbers should be used; if it is a rotation, integer quantum numbers should be used. Even though physicists in the early 1920s realized that one often obtained better results with half-integer quantum number  than with integer ones, they did not provide any systematic justification for using one rather than the other. In fact, the extra terms and factors of $\frac{1}{2}$ introduced in this period turned  out to come from a variety of sources, including the as yet undiscovered electron spin. It was only with the arrival of the new quantum theory that a systematic justification for these extra terms and factors could be given. As we saw in section 6, the WKB method  provided an important part of that justification. 

The other problem, the non-uniqueness of the quantized orbits, strikes right at the heart of the old quantum theory. Its residue in the new quantum theory is that the eigenstates found by separating and solving the Schr\"odinger equation in one set of coordinates can always be written as a superposition of the eigenstates found by doing so in another set of coordinates. This, of course, is no problem at all. But rather than showing how to solve the non-uniqueness problem in the old quantum theory, this evaporation of the problem in the new quantum theory merely shows that, contrary to one of the main articles of faith of the old quantum theory, orbits cannot be used to represent physical states in atomic physics.

\end{document}